\documentclass[aip,amsmath,amssymb,pop,graphicx]{revtex4-1}

\usepackage{graphicx}% Include figure files
\usepackage{dcolumn}% Align table columns on decimal point
\usepackage{bm}% bold math
\usepackage[utf8]{inputenc}
\usepackage[T1]{fontenc}
\usepackage{mathptmx}
\usepackage{etoolbox}

\draft % marks overfull lines with a black rule on the right

\begin{document}

%\preprint{}

\title{Theory of wakefield in a transversely inhomogeneous plasma waveguide} %Title of paper



\author{K. V. Galaydych}
\email[]{kgalaydych@gmail.com}
\author{P.I. Markov}
\author{G.V. Sotnikov}
%\thanks{}
\affiliation{National Science Center Kharkiv Institute of Physics and Technology 1, Akademichna St., Kharkiv, 61108, Ukraine}


%\date{\today}

\begin{abstract}
Theoretical studies have been made into the relativistic drive bunch generation of a wakefield in a cylindrical waveguide filled with a transversely inhomogeneous plasma. According to the model used, the transversely inhomogeneous plasma is considered as a combination of tubular plasma and the plasma background of different density. Analytical expressions have been derived for the excited radial and axial electric field components, and for the azimuthal magnetic field component. The dispersion of the plasma waveguide under study, as well as the topography of the electromagnetic field components of the TM-eigenwaves, resonant with the bunch, have been investigated. Longitudinal and transverse amplitude distribution structures of the axial and radial wakefields have been determined. Spectrum analysis of the longitudinal and transverse wakefields has been performed with the result that their frequency content has been determined.
\end{abstract}


\maketitle %\maketitle must follow title, authors, abstract and \pacs

\section{Introduction}
%\label{}
%\subsection{}
%\subsubsection{}
Charged particle accelerators are among the basic tools for pursuing fundamental and applied investigations. The research into the substance structure on the progressively smaller spatial scales, and the studies on the processes occurring within progressively shorter periods of time call for further increase of charged particle energy. In turn, increasing the particle energy with the use of traditional acceleration structures and the methods of their excitation, leads to an increase in the size of the design  accelerator complexes. On attaining limit values of electromagnetic field intensity amplitudes (of about tens of MV/m) in classical charged particle accelerators~\cite{Sessler1990,Dobert,Wuensch:IPAC2017-MOYCA1,Dolgashev2021}, investigations are now being actively conducted into new methods of acceleration, despite the challenges arising during their development and implementation~\cite{EuropeanStrategy,Colby2016,Cros2017,Albert2021,Assmann2020,Joshi2020}. These methods, among which we mention plasma wakefield acceleration~\cite{Chen1985}, laser wakefield acceleration~\cite{Tajima1979}, dielectric wakefield acceleration~\cite{Keinigs}, dielectric laser acceleration~\cite{Palmer}, seem attractive and promising, particularly, due to the possibility of attaining higher acceleration field amplitudes (of order of tens GV/m)~\cite{Faure2006,Gonsalves2019,Thompson2008,O'Shea2016}, i.e., being higher than in the conventional accelerator systems.  The successful realization of these methods can lead to the creation of future accelerators being much smaller in size, as well as more compact future free electron lasers, and charged particle colliders~\cite{Gruner2007,Leemans2009,Schulte2016,Joshi2019,Adli2022,Albert2023}. Reviews~\cite{Hogan2016,Nakajima2016,Adli2016,Wootton2016,Jing2016} describe the state-of-the-art of the studies on the above-mentioned new acceleration methods.

Among various problems of microwave electronics that are now being investigated analytically, numerically and experimentally, the main challenge on the way of implementing new methods of charged particle acceleration is to devise a technique for generating an intense wakefield. The mechanism for the wakefield excitation in accelerating waveguides is the Cherenkov emission. For efficient electromagnetic field generation by this mechanism with the use of drive bunches, and for the following efficient test bunch acceleration in this field, the waveguides must maintain the propagation of eigenwaves having the nonzero longitudinal electric field component and the phase velocity lower than the speed of light. The acceleration structures that suited as possible candidates for potential accelerating systems are the dielectric waveguides of cylindrical and rectangular configurations~\cite{Jing2016}. Generally, the channels for charged particle transportation are vacuum. In Ref.~\cite{Sotnikov2014,GALAYDYCH2024,Sotnikov2020,Markov2022}, consideration has been given to the wakefield excitation by the relativistic electron bunch in a cylindrical plasma-dielectric waveguide. The authors of the papers~\cite{Sotnikov2014,GALAYDYCH2024} have analytically demonstrated the possibility of radially stable acceleration of electron and positron test bunches in the excited wakefield in such an accelerating structure, which combines the properties of the dielectric electrodynamic system and the plasma waveguide, or the hollow plasma channel. The conclusions of the analytical theories have been confirmed by the numerical simulation results~\cite{Sotnikov2020,Markov2022}. The wakefield investigations in the cylindrical plasma-dielectric acceleration structure~\cite{GALAYDYCH2022,SOTNIKOV2025170522} have shown that in contrast to the cylindrical dielectric waveguide with no plasma filling, the presence of the initial bunch displacement does not lead to the BBU. In this way it has been demonstrated that the plasma in the charged particle transportation channel has a stabilizing effect on the BBU instability.

Along with dielectric waveguides, there exist other structures, which can maintain propagation of slow waves in them. One of these electrodynamic structures is the cylindrical metal waveguide filled with transversely inhomogeneous (along the radial distance from the waveguide axis)  plasma~\cite{Kondratenko,STAFFORD2001,Markov2008,Kartashov2022}. It is proposed in the present work that instead of a dielectric-homogeneous plasma combination, a transversely inhomogeneous plasma-filled waveguide should be used for generating in it (through the use of the relativistic driver bunch) the intense wakefield. That field would make possible the radially stable acceleration of both the electron and positron test bunches. As regards the plasma filling, the idea consists in forming the plasma filling in the waveguide in such a way that it would have a clearly defined region of tubular plasma and the plasma background of different densities. The plasma filling configuration of this type can be created using the laser beams, which are described by high-order Bessel functions~\cite{Fan2000,Gessner2016}. With this plasma density combination one may expect the realization of the situation, where the higher-density plasma (tubular plasma) would be responsible for the test-bunch acceleration, while the lower-density plasma (plasma background) would be responsible for the bunch focusing. Note that to make the management of the process easier, the wavelengths of the accelerating wave and the focusing wave must be essentially different.

It should be mentioned that the statements of problems similar to that under discussion have been considered earlier in Refs.~\cite{Balakirev1997,Pukhov2018}. The authors of paper~\cite{Balakirev1997} have analytically examined the linear regime of wakefield excitation by a relativistic electron bunch in a transversely inhomogeneous plasma for the cases, where the transverse boundedness of the accelerating structure (metal coating) is absent. Namely, consideration has been given to wakefield generation in a homogeneous plasma cylinder surrounded by unbounded vacuum space, and also, to the wakefield generation in a vacuum cylindrical channel formed in unbounded plasma.  At that, the bunch and plasma parameters were determined, at which the wakefields might be excited most efficiently. In turn, the radial accelerating structure boundedness causes changes in the dispersion characteristics, and in the electromagnetic field topography of the eigenwaves, which can occur and propagate in the accelerating structure. Paper~\cite{Pukhov2018} has numerically explored the nonlinear regime of wakefield generation by the relativistic electron bunch in the hollow plasma channel with a coaxial plasma filament of small radius. The presence of the coaxial plasma filament in the hollow plasma channel has been demonstrated to lead to the BBU stabilization of the electron drive bunch, and to a possible radially stable test bunch acceleration. Through particle-in-cell simulations and analysis the obtained results a physical mechanism of this stabilization was figured out, which consists of scattering electrons from the filament by the driver, and further nonlinear ions dynamics. The proposed accelerating structure allowed the authors to obtain high transformer ratio, high efficiency and low energy spread of the test electron bunch in the case of ramped current profile of the electron driver.

Though the issue of plasma channels creation goes beyond the scope of the present theoretical work, it should be noted that the studies being pursued in this line of research are currently urgent and are of particular importance for the plasma methods of acceleration.  Among these methods, it might be well to point out the following:   heat-pipe oven~\cite{Muggli1999,Lishilin2019}, capillary discharges~\cite{Spence2000,Lopes2003,Bobrova2013,ManwaniIPAC2024}, hydrodynamic plasma channels (expansion waveguides)~\cite{Durfee1995,Sheng2005}, hydrodynamic optical-field-ionized plasma channels~\cite{Shalloo2018,Mewes2023} (more information about them can be found, for example, in review~\cite{garland2020plasmasourcesdiagnostics}).

In the present paper, we analyze, both analytically and numerically, the wakefield excitation by a  single relativistic electron bunch in a cylindrical waveguide filled with cold transversely inhomogeneous plasma. In the derivation of the expressions for the generated wakefield components, we simultaneously took into account two factors, which may be typical for the real experimental conditions, namely, (I) the transverse boundedness of the accelerating structure (metal coating), and (II) the radial plasma inhomogeneity (as a simplified model, which permits the construction of an analytical theory).

\section{Statement of the problem}
\begin{figure}
  \centering
  \includegraphics[width=0.7\textwidth]{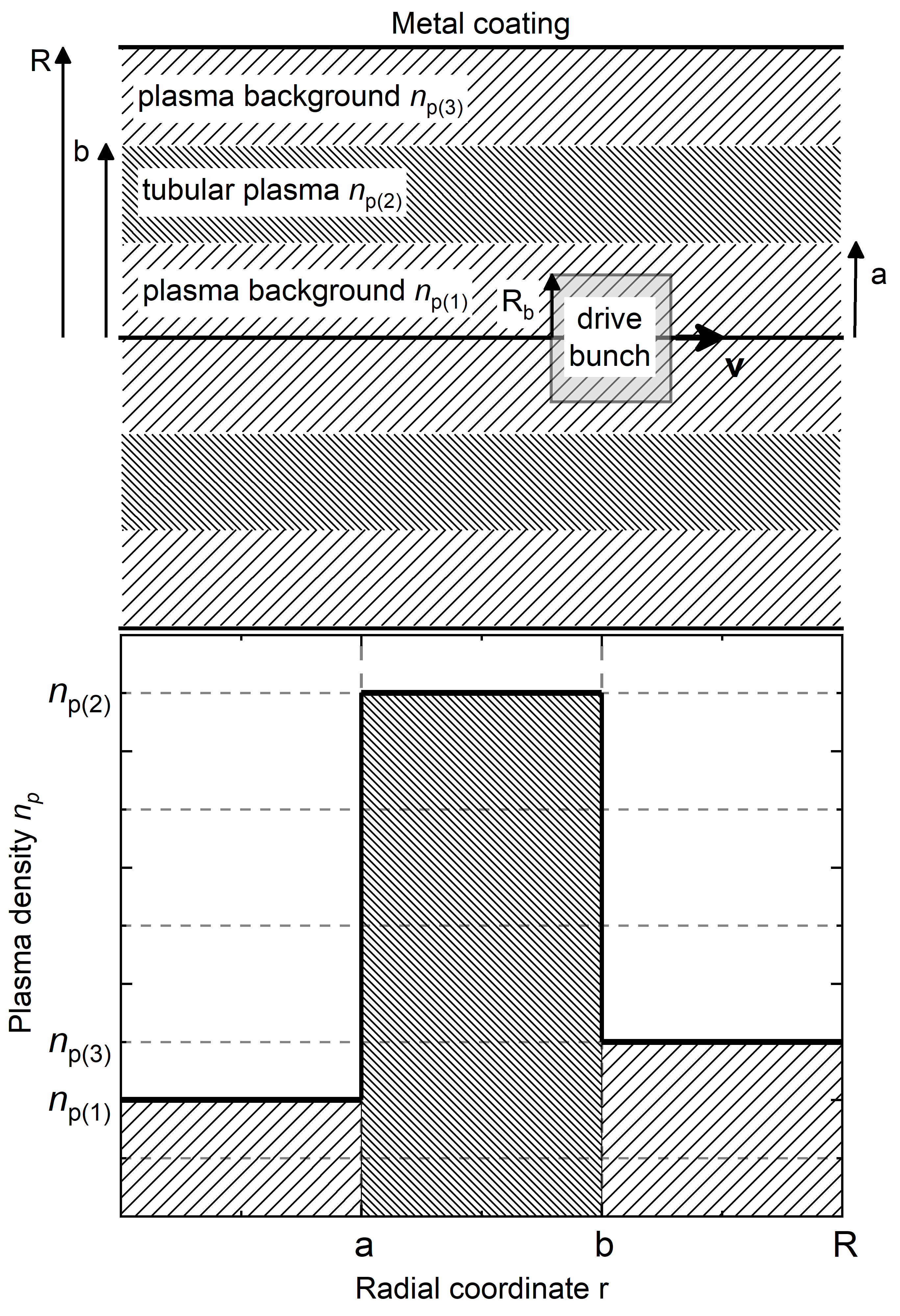}
  \caption{(Top) General view of the cylindrical plasma waveguide. Metal coating, transversely inhomogeneous plasma and drive bunch are shown schematically. The on-axis drive bunch moves along the waveguide axis, (bottom) The waveguide plasma density versus the radial coordinate is shown schematically.}\label{Fig:01}
\end{figure}

The accelerating structure under study presents a circular metal waveguide of radius $R$, which is partially filled with transversely inhomogeneous plasma (Fig.~\ref{Fig:01}). As a model transversely inhomogeneous plasma we consider a combination of three plasmas having the densities $n_{p(1)}$, $n_{p(2)}$ and $n_{p(3)}$. The plasmas of densities $n_{p(1)}$ and $n_{p(3)}$ make the plasma background.  The tubular plasma of higher density is represented by the plasma $n_{p(2)}$. In each of the partial regions (1), (2) and (3) the plasma is homogeneous in the transverse cross-section. In the plasma of density $n_{p(1)}$, the drive electron bunch of charge $Q_b$, length $L_b$ and radius $R_b$ propagates at a constant velocity $v$ along the waveguide axis.

In the construction of the analytical theory the following basic approximations and assumptions have been made. Here, we assume that the plasma is linear. This assumption will be verified in numerical calculations for particular parameters of the drive bunch and the plasma waveguide. It is also assumed that there is no evolution of the drive bunch in the process of wakefield generation in the waveguide.  We restrict the discussion to the case, where the drive bunch radius is smaller than the inner radius of the tubular plasma ($R_b < a$). The plasma is supposed to be cold, collisionless and fully ionized. The external magnetic field is absent.

The aim of this work has been to develop the analytical theory of wakefield generation by a driver bunch in a cylindrical waveguide filled with a transversely inhomogeneous plasma.  Since the electron bunch is axially symmetric, the generation of only azimuthally symmetric electromagnetic field components will be analyzed.
\section{Analytical studies}
To determine the wakefield excited in the waveguide by a drive bunch of arbitrary charge and current distribution, this the bunch may be conceived as a set of infinitely thin and short rings, and as a preliminary, we shall determine the wakefield of such an infinitely short and thin charged ring. The charge and current densities of the ring particle of radius $r_0$ and charge $q_0$, which moves at a constant velocity v along the waveguide axis ($z$ axis) without  transverse displacement, may be written as
\begin{equation}\label{eq:01}
\begin{split}
\rho(r,z,t) = \frac{q_0}{2\pi r_0v}\delta(r-r_0)\delta(t-t_0-z/v),\,j_z = v\rho,
\end{split}
\end{equation}
where $t_0$  is the time taken by the particle to pass across the coordinate $z=0$, and $\delta(x)$ is the Dirac delta function. By virtue of the fact that the problem is homogeneous in both the time and the direction parallel to the drive bunch velocity, it appears convenient to represent all the quantities as the Fourier double integral over the frequency $\omega$, and over the longitudinal component of the wavenumber $k_z$, parallel to the bunch vector velocity. Then we have
\begin{equation}\label{eq:02}
\begin{split}
\begin{Bmatrix}\mathbf{E}(r,z,t)\\\mathbf{D}(r,z,t)\\\mathbf{H}(r,z,t)\\\end{Bmatrix}&= \int\limits_{ - \infty }^{ + \infty }dk_z\int\limits_{ - \infty }^{ + \infty } {d\omega \begin{Bmatrix}\mathbf{E}_\omega (r,k_z,\omega )\\\varepsilon(\omega)\mathbf{E}_\omega (r,k_z,\omega )\\\mathbf{H}_\omega (r,k_z,\omega )\\\end{Bmatrix} }{e^{ ik_{z}z - i\omega t}}{\text{,}}\\
\begin{Bmatrix}\rho(r,z,t)\\j_z(r,z,t)\\\end{Bmatrix}&= \int\limits_{ - \infty }^{ + \infty }dk_z\int\limits_{ - \infty }^{ + \infty } {d\omega \begin{Bmatrix}\rho_\omega (r,k_z,\omega )\\j_{z\omega} (r,k_z,\omega )\\\end{Bmatrix} }{e^{ik_{z}z - i\omega t}}{\text{.}}
\end{split}
\end{equation}
The dielectric permittivity $\varepsilon(\omega)$ in each of the plasma homogeneity regions $i=(1)-(3)$ is determined by the corresponding plasma density $n_{p(i})$
\begin{equation}\label{eq:03}
\varepsilon(\omega) = \varepsilon_{p(i)}(\omega) =  1 - \omega _{p(i)}^2/{\omega ^2}, \quad \omega^2_{p(i)}=4\pi e^2n_{p(i)}/m_e
\end{equation}
where $e$ and $m_e$ denotes the electron charge and mass, respectively, the ion contribution to the dielectric permittivity will be neglected. From using Maxwell’s equations for the wakefield Fourier-components we obtain the wave equation for the axial component of the electric field $E_{z\omega}$ in the $i$-th region of the piecewise homogeneous plasma
\begin{equation}\label{eq:04}
\begin{split}
\frac{1}{r}\frac{\partial }{{\partial r}}\left( {r\frac{{\partial E_{z\omega} }}{{\partial r}}} \right) + k^2_{\perp p(i)}E_{z\omega} = \frac{iq_0\delta (r-r_0)\delta (k_z - \omega /v)}{\pi r_0v\varepsilon_{p(i)}}\left( k_ze^{ik_zvt_0} - \frac{\varepsilon_{p(i)}\omega v}{c^2}e^{i\omega t_0} \right),
\end{split}
\end{equation}
where the transverse component of the wave vector $k^2_{\perp p(i)}$ is equal to $k^2_{\perp p(i)}=\varepsilon_{p(i)}\omega^2/c^2-k_{z}^2$. The Fourier components of the radial electric field $E_{r\omega}$ and the azimuthal magnetic field $H_{\varphi \omega}$ are expressed in terms of  $E_{z\omega}$
\begin{equation}\label{eq:05}
\begin{split}
H_{\varphi \omega} = \frac{\omega \varepsilon (\omega)}{ck_z}E_{r\omega},\,\,E_{r\omega} = \frac{ik_z}{\varepsilon (\omega)\omega^2/c^2 - k_z^2}\frac{\partial E_{z\omega}}{\partial r}.
\end{split}
\end{equation}
To determine the excited wakefield, by convention the waveguide is divided into four regions: (I)--$(0 \leq r \leq r_0)$, (II)--$(r_0 < r \leq a)$, (III)--$(a < r \leq b)$ and (IV)--$(b < r \leq R)$. The solutions to Eq.~(\ref{eq:04}) for each of the mentioned regions can be written as:
\begin{equation}\label{eq:06}
\begin{split}
&E_{z\omega}^{(I)} = A_{1}\frac{J_0(k_{\perp p(1)}r)}{J_0(k_{\perp p(1)}r_0)},\\
&E_{z\omega}^{(II)} = A_{1}\frac{F_0(k_{\perp p(1)}r,k_{\perp p(1)}a)}{F_0(k_{\perp p(1)}r_0,k_{\perp p(1)}a)} + A_{2}\frac{F_0(k_{\perp p(1)}r,k_{\perp p(1)}r_0)}{F_0(k_{\perp p(1)}a,k_{\perp p(1)}r_0)},\\
&E_{z\omega}^{(III)} = A_{2}\frac{F_0(k_{\perp p(2)}r,k_{\perp p(2)}b)}{F_0(k_{\perp p(2)}a,k_{\perp p(2)}b)} + A_{3}\frac{F_0(k_{\perp p(2)}r,k_{\perp p(2)}a)}{F_0(k_{\perp p(2)}b,k_{\perp p(2)}a)},\\
&E_{z\omega}^{(IV)} = A_{3}\frac{F_0(k_{\perp p(3)}r,k_{\perp p(3)}R)}{F_0(k_{\perp p(3)}b,k_{\perp p(3)}R)},
\end{split}
\end{equation}
where ${F_0}(x,y) = J_0(x)Y_0(y) - Y_0(x)J_0(y)$, $J_0$ and $Y_0$ are, the zeroth-order Bessel and Weber cylinder functions. With this form of writing the solutions to Eq.~(\ref{eq:04}), the axial component of the electric field $E_{z\omega}$ is continuous at the boundaries of plasma regions, while on the metal coating surface ($r=R$) vanishes. The constants $A_{1}$, $A_{2}$ and $A_{3}$ should be determined through the use of the remaining boundary conditions: the axial component of the magnetic field is continuous at the plasma region boundaries ($r=a,b$), while at the beam surface ($r=r_0$) this component experiences a current density-defined jump
\begin{equation}\label{eq:07}
\left( \frac{dE^{(II)}_{z\omega}}{dr} \right)_{r=r_0} - \left( \frac{dE^{(I)}_{z\omega}}{dr} \right)_{r=r_0} = \frac{iq_0\delta (k_z - \omega /v)}{\pi r_0v\varepsilon_{p(1)}}\left( k_ze^{ik_zvt_0} - \frac{\varepsilon_{p(1)}\omega v}{c^2}e^{i\omega t_0} \right).
\end{equation}
Substituting the first two expressions from system~(\ref{eq:06}) into Eq.~(\ref{eq:07}), and performing integration over the longitudinal component of the wavenumber $k_z$, we obtain the expressions for $E_{z\omega}$ in all four regions of the waveguide
\begin{equation}\label{eq:08}
\begin{split}
E^{(I)}_{z\omega} = -\frac{iq_0}{\pi a\omega}\frac{J_0(\varkappa_{p(1)}r)J_0(\varkappa_{p(1)}r_0)}{J^2_0(\varkappa_{p(1)}a)}\frac{D_{2}(\omega ,\omega/v)}{D(\omega ,\omega/v)}e^{i\omega t_0 + i\omega/vz} +\\
\frac{iq_0\varkappa^2_{p(1)}}{2\omega \varepsilon_{p(1)}}\frac{J_0(\varkappa_{p(1)}r)F_0(\varkappa_{p(1)}r_0,\varkappa_{p(1)}a)}{J_0(\varkappa_{p(1)}a)}e^{i\omega t_0 + i\omega/vz},
\end{split}
\end{equation}
\begin{equation}\label{eq:09}
\begin{split}
E^{(II)}_{z\omega} = -\frac{iq_0}{\pi a\omega}\frac{J_0(\varkappa_{p(1)}r)J_0(\varkappa_{p(1)}r_0)}{J^2_0(\varkappa_{p(1)}a)}\frac{D_{2}(\omega ,\omega/v)}{D(\omega ,\omega/v)}e^{i\omega t_0 + i\omega/vz} +\\
\frac{iq_0\varkappa^2_{p(1)}}{2\omega \varepsilon_{p(1)}}\frac{J_0(\varkappa_{p(1)}r_0)F_0(\varkappa_{p(1)}r,\varkappa_{p(1)}a)}{J_0(\varkappa_{p(1)}a)}e^{i\omega t_0 + i\omega/vz},
\end{split}
\end{equation}
\begin{equation}\label{eq:10}
\begin{split}
E^{(III)}_{z\omega} = -\frac{iq_0}{\pi a\omega D(\omega ,\omega/v)}\frac{J_0(\varkappa_{p(1)}r_0)}{J_0(\varkappa_{p(1)}a)}\left( \frac{F_0(\varkappa_{p(2)}r,\varkappa_{p(2)}b)}{F_0(\varkappa_{p(2)}a,\varkappa_{p(2)}b)}D_2(\omega ,\omega/v) + \right. \\ \left.\frac{F_0(\varkappa_{p(2)}r,\varkappa_{p(2)}a)}{F_0(\varkappa_{p(2)}b,\varkappa_{p(2)}a)}\frac{F_1(\varkappa_{p(2)}b,\varkappa_{p(2)}b)}{\varkappa_{p(2)}F_0(\varkappa_{p(2)}a,\varkappa_{p(2)}b)} \right)e^{i\omega t_0 + i\omega/vz},
\end{split}
\end{equation}
\begin{equation}\label{eq:11}
\begin{split}
E^{(IV)}_{z\omega} = -\frac{iq_0}{\pi \varkappa_{p(2)}a\omega D(\omega ,\omega/v)}\frac{J_0(\varkappa_{p(1)}r_0)}{J_0(\varkappa_{p(1)}a)}\frac{F_1(\varkappa_{p(2)}b,\varkappa_{p(2)}b)}{F_0(\varkappa_{p(2)}a,\varkappa_{p(2)}b)}\times\\
\frac{F_0(\varkappa_{p(3)}r,\varkappa_{p(3)}R)}{F_0(\varkappa_{p(3)}b,\varkappa_{p(3)}R)}e^{i\omega t_0 + i\omega/vz},
\end{split}
\end{equation}
where $\varkappa_{p(1)} = k_{\perp p(1)}(k_z=\omega/v)$,\,$\varkappa_{p(2)} = k_{\perp {p(2)}}(k_z=\omega/v)$,\,$\varkappa_{p(3)} = k_{\perp p(3)}(k_z=\omega/v)$, $F_1(x,y) = -J_1(x)Y_0(y)+Y_1(x)J_0(y)$, $J_1$ and $Y_1$ are the Bessel and Weber cylindrical first-order functions. In Eqs.~(\ref{eq:08})--(\ref{eq:11}) the expression for $D(\omega ,k_z)$ is defined as
\begin{equation}\label{eq:12}
\begin{split}
D(\omega ,k_z) &= D_1(\omega ,k_z)D_2(\omega ,k_z) + D_3(\omega ,k_z),\\
D_1(\omega ,k_z) &= \frac{\varepsilon_{p(1)}}{k_{\perp p(1)}}\frac{J_1(k_{\perp p(1)}a)}{J_0(k_{\perp p(1)}a)} + \frac{\varepsilon_{p(2)}}{k_{\perp p(2)}}\frac{F_1(k_{\perp p(2)}a,k_{\perp p(2)}b)}{F_0(k_{\perp p(2)}a,k_{\perp p(2)}b)},\\
D_2(\omega ,k_z) &= \frac{\varepsilon_{p(3)}}{k_{\perp p(3)}}\frac{F_1(k_{\perp p(3)}b,k_{\perp p(3)}R)}{F_0(k_{\perp p(3)}b,k_{\perp p(3)}R)} - \frac{\varepsilon_{p(2)}}{k_{\perp p(2)}}\frac{F_1(k_{\perp p(2)}b,k_{\perp p(2)}a)}{F_0(k_{\perp p(2)}b,k_{\perp p(2)}a)},\\
D_3(\omega ,k_z) &= \frac{\varepsilon_{p(2)}^2}{k_{\perp p(2)}^2}\frac{F_1(k_{\perp p(2)}a,k_{\perp p(2)}a)F_1(k_{\perp p(2)}b,k_{\perp p(2)}b)}{F_0(k_{\perp p(2)}b,k_{\perp p(2)}a)F_0(k_{\perp p(2)}a,k_{\perp p(2)}b)}.
\end{split}
\end{equation}
Having made the Fourier inverse frequency $\omega$ transform from expressions Eqs.~(\ref{eq:08})--(\ref{eq:11}) we obtain the equations for the longitudinal component of the electric field. The integration over frequency reduces to the summation over the residues in the poles of the integrands, which are defined by the zeros of the functions $\varepsilon_{p(1)}(\omega)$ and $D(\omega ,\omega/v)$. The equation  $\varepsilon_{p(1)}(\omega)=0$ determines the frequency of the plasma (Langmuir) wave $\omega_{p(1)}$ in the region $r\leq a$, and the corresponding poles are found only in the expressions for the Fourier components of the electric field in this region. The equation $D(\omega ,\omega/v)=0$ defines the waveguide eigenfrequencies $\omega_s$, being resonant with the drive bunch (Cherenkov resonance), and the corresponding poles are present in the expressions for the Fourier components of both the electric and magnetic fields throughout the waveguide regions. Finally, the longitudinal electric field, excited by the ring charged particle, has the form:
\begin{equation}\label{eq:13}
\begin{split}
E^{(I)}_z(r,z,t,r_0,t_0) = 2q_0k^2_{p(1)}\frac{I_0(k_{p(1)}r)}{I_0(k_{p(1)}a)}\Delta_{0}(k_{p(1)}r_0,k_{p(1)}a)\times\\
\cos\omega_{p(1)}(t-t_0-z/v)\theta(t-t_0-z/v) - \sum_{s}^{}\frac{4q_0}{a\omega_s}\frac{J_0(\varkappa_{p(1)s}r)J_0(\varkappa_{p(1)s}r_0)}{J^2_0(\varkappa_{p(1)s}a)}\frac{D_{2s}}{D'({\omega_s})}\times\\
\cos\omega_s(t-t_0-z/v)\theta(t-t_0-z/v),
\end{split}
\end{equation}
\begin{equation}\label{eq:14}
\begin{split}
E^{(II)}_z(r,z,t,r_0,t_0) = 2q_0k^2_{p(1)}\frac{I_0(k_{p(1)}r_0)}{I_0(k_{p(1)}a)}\Delta_{0}(k_{p(1)}r,k_{p(1)}a)\times\\
\cos\omega_{p(1)}(t-t_0-z/v)\theta(t-t_0-z/v) - \sum_{s}^{}\frac{4q_0}{a\omega_s}\frac{J_0(\varkappa_{p(1)s}r)J_0(\varkappa_{p(1)s}r_0)}{J^2_0(\varkappa_{p(1)s}a)}\frac{D_{2s}}{D'({\omega_s})}\times\\
\cos\omega_s(t-t_0-z/v)\theta(t-t_0-z/v),
\end{split}
\end{equation}
\begin{equation}\label{eq:15}
\begin{split}
E^{(III)}_z(r,z,t,r_0,t_0) = -\sum_{s}^{}\frac{4q_0}{a\omega_s D'({\omega_s})}\frac{J_0(\varkappa_{p(1)s}r_0)}{J_0(\varkappa_{p(1)s}a)}\left( \frac{F_0(\varkappa_{p(2)s}r,\varkappa_{p(2)s}b)}{F_0(\varkappa_{p(2)s}a,\varkappa_{p(2)s}b)}D_{2s} + \right. \\ \left.\frac{F_0(\varkappa_{p(2)s}r,\varkappa_{p(2)s}a)}{F_0(\varkappa_{p(2)s}b,\varkappa_{p(2)s}a)}\frac{\varepsilon_{p(2)s}F_1(\varkappa_{p(2)s}b,\varkappa_{p(2)s}b)}{\varkappa_{p(2)s}F_0(\varkappa_{p(2)s}a,\varkappa_{p(2)s}b)} \right)\cos\omega_s(t-t_0-z/v)\theta(t-t_0-z/v),
\end{split}
\end{equation}
\begin{equation}\label{eq:16}
\begin{split}
E^{(IV)}_z(r,z,t,r_0,t_0) = -\sum_{s}^{}\frac{4q_0}{a\omega_s D'({\omega_s})}\frac{J_0(\varkappa_{p(1)s}r_0)}{J_0(\varkappa_{p(1)s}a)}\frac{\varepsilon_{p(2)s}F_1(\varkappa_{p(2)s}b,\varkappa_{p(2)s}b)}{\varkappa_{p(2)s}F_0(\varkappa_{p(2)s}a,\varkappa_{p(2)s}b)}\times\\
\frac{F_0(\varkappa_{p(3)s}r,\varkappa_{p(3)s}R)}{F_0(\varkappa_{p(3)s}b,\varkappa_{p(3)s}R)}\cos\omega_s(t-t_0-z/v)\theta(t-t_0-z/v),
\end{split}
\end{equation}
where $k_{p(1)}=\omega_{p(1)}/v$, $\varkappa_{p(1)s} = \varkappa_{p(1)}(\omega_s)$,\,$\varkappa_{p(2)s} = \varkappa_{p(2)}(\omega_s)$,\,$\varkappa_{p(3)s} = \varkappa_{p(3)}(\omega_s)$,\,$D_{2s}=D_{2}(\omega_s ,\omega_s/v)$, and $D'(\omega_s)=\left( \frac{d D(\omega ,\omega/v)}{d \omega} \right)_{\omega=\omega_s}$, $\Delta _0(x,y) = I_0(x)K_0(y) - K_0(x)I_0(y)$, $I_{0}$ and $K_{0}$ are the modified zeroth-order Bessel and McDonald  functions.
Expressions~(\ref{eq:13})--(\ref{eq:16}) demonstrate that the longitudinal component of the electric field in the plasma background region at $r\leq a$ is contributed by two summands, namely, the wakefield of the Langmuir wave at frequency $\omega_{p(1)}$, and the wakefield of the eigenwaves, resonant with the drive bunch at frequencies $\omega_{s}$. Apparently, the wakefield of the Langmuir wave is localized in the plasma background region at $r\leq a$,  whereas in the tubular plasma regions and in the plasma background region at $b \leq r\leq a$, it is absent.  Physically, this is due to the Langmuir wave field screening by the surface charge induced at the plasma background-tubular plasma interface~\cite{Balakirev1997}.

To determine wakefield excited by the bunch of finite size, there is a need to summate the wakefields generated by the whole set of infinitely thin and short  charged rings, in the form of which  the drive bunch has been initially represented. We assume that the drive bunch has the following charge density distribution within its boundaries: the rectangular profile in the radial direction and the cosine-like profile in the longitudinal direction. In going from the ring-shaped charged particle to the bunch of finite sizes in the  longitudinal and transverse directions, it is necessary that in expressions~(\ref{eq:13})--(\ref{eq:16}) the charge $q_0$  should be replaced by the expression $q_0 = (Q_b/\pi R^2_bT_b)(1-\cos2\pi t_0/T_b)2\pi dt_0dr_0r_0$, and be integrated over the radius $r_0$ in the range from  0 to $R_b$, and also, over the time of fly into the waveguide, $t_0$, within the range  0 to $T_b$, where $T_b = L_b/v$ is the time duration of the drive bunch.

The final expression for the axial wakefield excited by the finite-size drive bunch, in the plasma background region at $r\leq a$ have the following form:
\begin{equation}\label{eq:17}
\begin{split}
E_z(r,z,t) = -\frac{4Q_b}{R_bL_b}\left( [\theta(r)-\theta(r-R_b)]R_{z1}(r) + \right. \\ \left.
[\theta(r-R_b)-\theta(r-a)]R_{z2}(r) \right)\Psi_{\parallel}(t,z,\omega_{p(1)}) -\\
\sum_{s}^{}\frac{8Q_bv}{R_bL_ba\varkappa_{p(1)s}\omega^2_s}\frac{D_{2s}}{D'({\omega_s})}
\frac{J_1(\varkappa_{p(1)s}R_b)J_0(\varkappa_{p(1)s}r)}{J_0^2(\varkappa_{p(1)s}a)}\Psi_{\parallel}(t,z,\omega_s),
\end{split}
\end{equation}

\begin{equation}\label{eq:18}
\begin{split}
&R_{z1}(r) = \frac{1}{k_{p(1)}R_b} - \frac{I_0(k_{p(1)}r)}{I_0(k_{p(1)}a)}\Delta_{1}(k_{p(1)}R_b,k_{p(1)}a),\\
&R_{z2}(r) = -\frac{I_1(k_{p(1)}R_b)}{I_0(k_{p(1)}a)}\Delta_{0}(k_{p(1)}r,k_{p(1)}a),
\end{split}
\end{equation}

\begin{equation}\label{eq:19}
\begin{split}
E_z(r,z,t) = -\sum_{s}^{}\frac{8Q_bv}{R_bL_ba\varkappa_{p(1)s}\omega_s D'({\omega_s})}\frac{J_1(\varkappa_{p(1)s}R_b)}{J_0(\varkappa_{p(1)s}a)}\left( \frac{F_0(\varkappa_{p(2)s}r,\varkappa_{p(2)s}b)}{F_0(\varkappa_{p(2)s}a,\varkappa_{p(2)s}b)}D_{2s} + \right. \\ \left.\frac{F_0(\varkappa_{p(2)s}r,\varkappa_{p(2)s}a)}{F_0(\varkappa_{p(2)s}b,\varkappa_{p(2)s}a)}\frac{\varepsilon_{p(2)s}F_1(\varkappa_{p(2)s}b,\varkappa_{p(2)s}b)}{\varkappa_{p(2)s}F_0(\varkappa_{p(2)s}a,\varkappa_{p(2)s}b)} \right)\Psi_{\parallel}(t,z,\omega_s),
\end{split}
\end{equation}

\begin{equation}\label{eq:20}
\begin{split}
E_z(r,z,t) = -\sum_{s}^{}\frac{8Q_bv}{R_bL_ba\varkappa_{p(1)s}\omega_s D'({\omega_s})}\frac{J_1(\varkappa_{p(1)s}R_b)}{J_0(\varkappa_{p(1)s}a)}\frac{\varepsilon_{p(2)s}F_1(\varkappa_{p(2)s}b,\varkappa_{p(2)s}b)}{\varkappa_{p(2)s}F_0(\varkappa_{p(2)s}a,\varkappa_{p(2)s}b)}\times\\
\frac{F_0(\varkappa_{p(3)s}r,\varkappa_{p(3)s}R)}{F_0(\varkappa_{p(3)s}b,\varkappa_{p(3)s}R)}\Psi_{\parallel}(t,z,\omega_s).
\end{split}
\end{equation}
The corresponding expressions for the transverse wakefield components in the plasma background region at $r\leq a$ are written as:
\begin{equation}\label{eq:21}
\begin{split}
E_r(r,z,t) = -\frac{4Q_b}{R_bL_b}\left( [\theta(r)-\theta(r-R_b)]R_{r1}(r) + \right. \\ \left.
[\theta(r-R_b)-\theta(r-a)]R_{r2}(r) \right)\Psi_{\perp}(t,z,\omega_{p(1)}) +\\
\sum_{s}^{}\frac{8Q_bv}{R_bL_ba\varkappa^2_{p(1)s}\omega_s a}\frac{D_{2s}}{D'({\omega_s})}
\frac{J_1(\varkappa_{p(1)s}R_b)J_1(\varkappa_{p(1)s}r)}{J_0^2(\varkappa_{p(1)s}a)}\Psi_{\perp}(t,z,\omega_s),
\end{split}
\end{equation}
\begin{equation}\label{eq:22}
\begin{split}
&R_{r1}(r) = \frac{I_1(k_{p(1)}r)}{I_0(k_{p(1)}a)}\Delta_{1}(k_{p(1)}R_b,k_{p(1)}a),\\
&R_{r2}(r) = \frac{I_1(k_{p(1)}R_b)}{I_0(k_{p(1)}a)}\Delta_{1}(k_{p(1)}r,k_{p(1)}a),
\end{split}
\end{equation}
\begin{equation}\label{eq:23}
\begin{split}
H_\varphi(r,z,t) = \sum_{s}^{}\frac{8\beta \varepsilon_{p(1)s}Q_b}{R_bL_ba\varkappa^2_{p(1)s}\omega_s a}\frac{D_{2s}}{D'({\omega_s})}
\frac{J_1(\varkappa_{p(1)s}R_b)J_1(\varkappa_{p(1)s}r)}{J_0^2(\varkappa_{p(1)s}a)}\Psi_{\perp}(t,z,\omega_s),
\end{split}
\end{equation}
where $\Delta _1(x,y) = I_1(x)K_0(y) + K_1(x)I_0(y)$, $I_{1}$ and $K_{1}$ are the modified first-order Bessel and McDonald functions. The longitudinal-temporal structure of the excited wakefield is described by the functions $\Psi_{\parallel}(t,z,\omega)$ and $\Psi_{\perp}(t,z,\omega)$, which have the form:
\begin{equation}\label{eq:24}
\begin{split}
\Psi_{\parallel}(t,z,\omega) = &\frac{1}{\omega T_b}\bigg( \theta(t - z/v)\sin\omega(t - z/v) - \theta(t - z/v - T_b)\sin\omega(t - z/v - T_b) \bigg) -\\
&\frac{1}{4\pi^2 - \omega^2 T_b^2}\left( \theta(t - z/v)\left\{2\pi \sin\frac{2\pi}{T_b}(t - z/v) - \omega T_b\sin\omega(t - z/v)\right\} -\right. \\& \left. \theta(t - z/v - T_b)\left\{2\pi \sin\frac{2\pi}{T_b}(t - z/v-T_b) - \omega T_b\sin\omega(t - z/v-T_b)\right\} \right),
\end{split}
\end{equation}
\begin{equation}\label{eq:25}
\begin{split}
\Psi_{\perp}(t,z,\omega) = &\frac{1}{\omega T_b}\bigg( \theta(t - z/v)\Big(1-\cos\omega(t - z/v)\Big) - \theta(t - z/v - T_b)\Big(1-\cos\omega(t - z/v - T_b)\Big) \bigg) -\\
&\frac{\omega T_b}{\omega^2 T_b^2 - 4\pi^2}\left( \theta(t - z/v)\left\{\cos\frac{2\pi}{T_b}(t - z/v) - \cos\omega(t - z/v)\right\} -\right. \\& \left. \theta(t - z/v-T_b)\left\{\cos\frac{2\pi}{T_b}(t - z/v-T_b) - \cos\omega(t - z/v-T_b)\right\} \right).
\end{split}
\end{equation}
The first summands in the expressions for the axial and radial components of the excited electric field~(\ref{eq:17})--(\ref{eq:23}), which are responsible for the Langmuir wave contribution to the total wakefield, are dependent only on the parameters of the drive bunch and the plasma background in the area of the waveguide axis. The second summands in the expressions for the axial and radial components of the excited electric field, which are responsible for the contribution of surface eigenwaves (resonant with the drive bunch) to the total wakefield, are dependent on the parameters of both the drive bunch and the plasma waveguide. Between the solutions of the Maxwell equations for the drive bunch-generated electric and magnetic fields~(\ref{eq:17})--(\ref{eq:23}), there is an important qualitative difference. The azimuthal component of the generated magnetic field~(\ref{eq:23}) is determined only by the field of surface eigenwaves. The Langmuir wave cannot contribute to the generated magnetic field, as the field of this wave is electrostatic in its nature, in contrast to the field of the eigenwaves, being in resonance with the drive bunch, which is electromagnetic in character.
\section{Numerical analysis}
\begin{figure}
  \centering
  \includegraphics[width=0.49\textwidth]{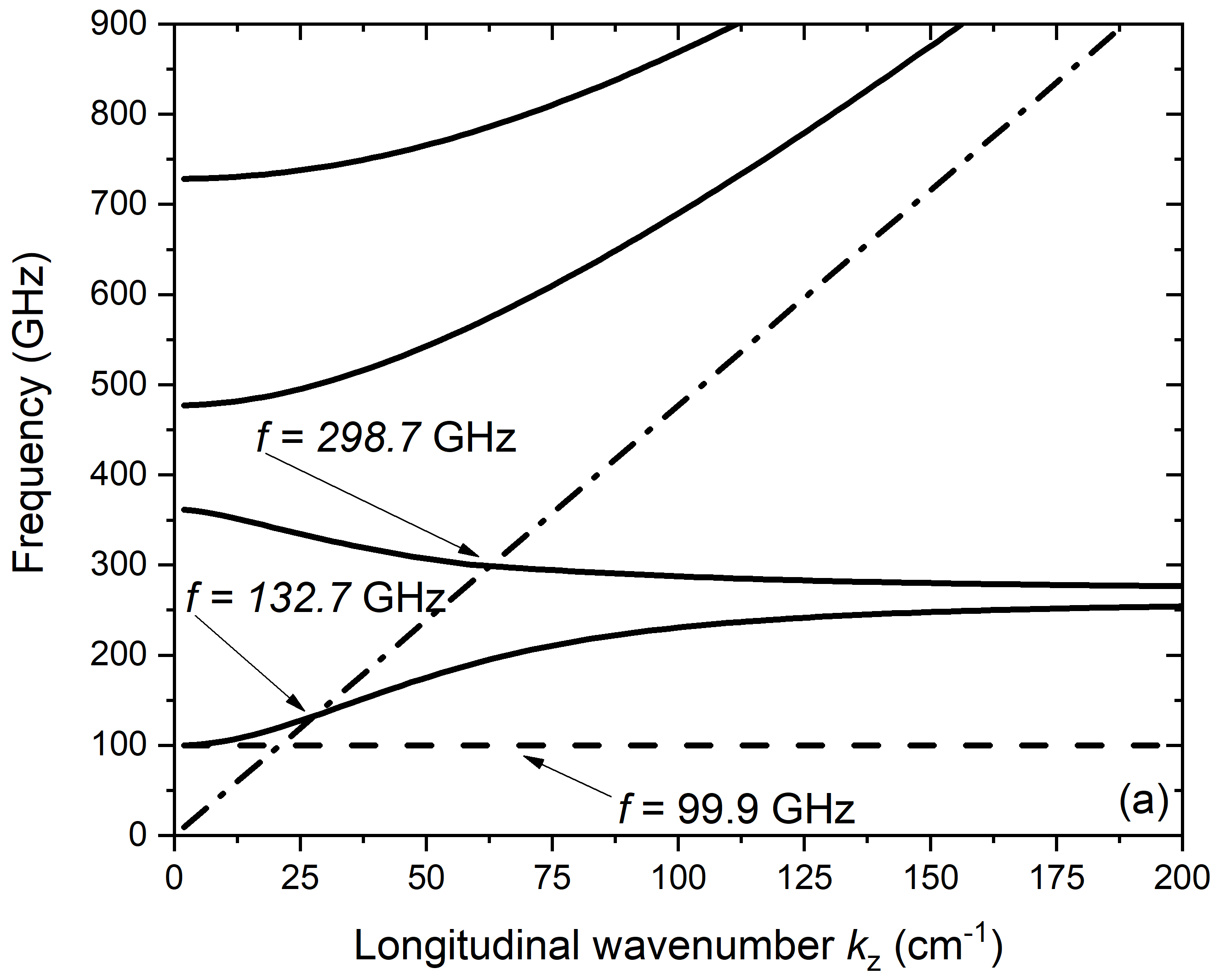}
  \includegraphics[width=0.49\textwidth]{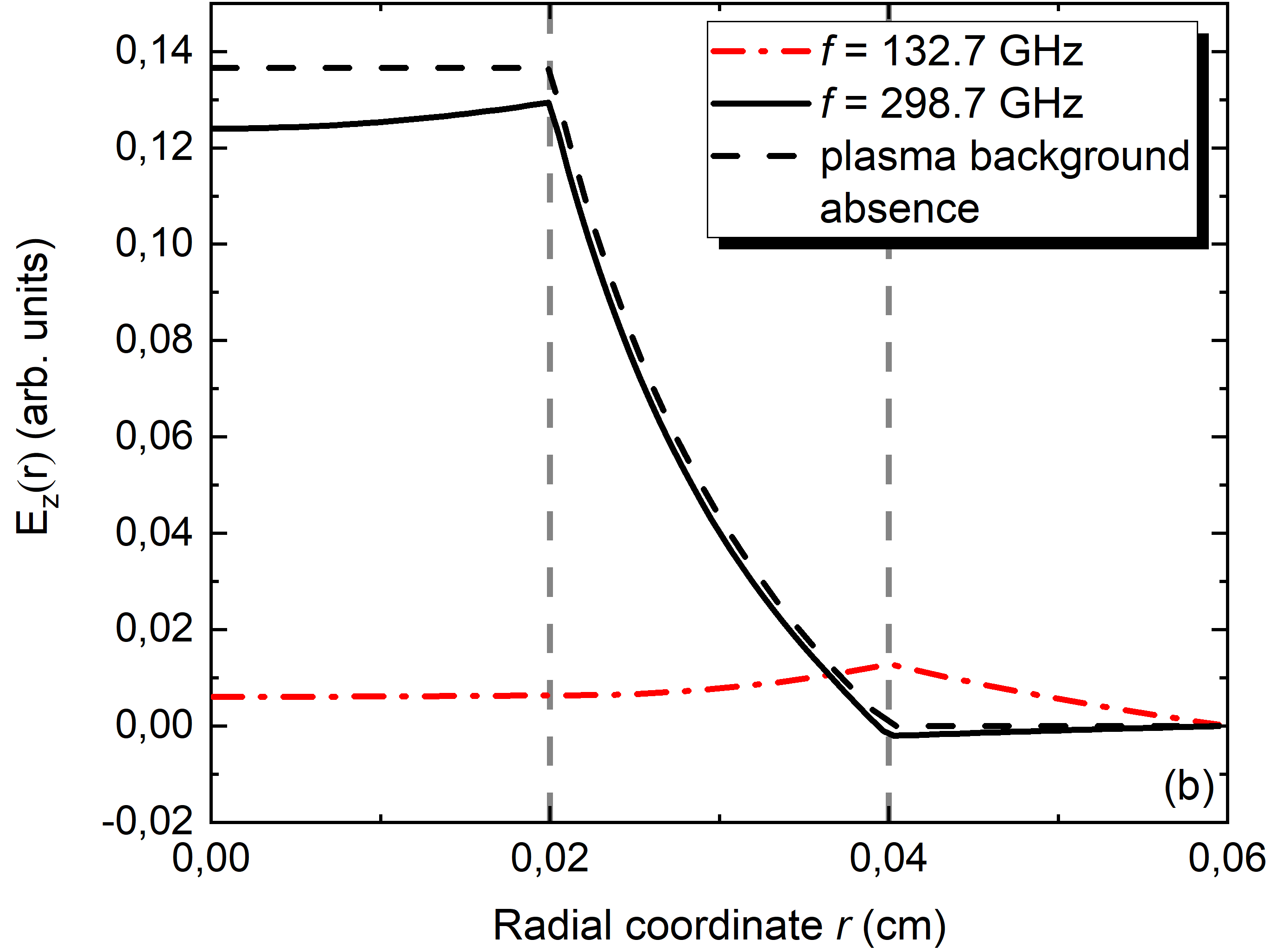}
  \caption{(a) Dispersion of the first four TM eigenwaves of the plasma waveguide, the dash-dotted line represents a beam line $\omega = k_{z}v$, (b) Topography of the waveguide eigenmodes resonant with the drive bunch. In addition, the resonant mode topography of the high-frequency surface wave is shown for the case with no plasma background.}\label{Fig:02}
\end{figure}

With analytical expressions~(\ref{eq:17})--(\ref{eq:25}) we numerically investigate the space-time structure of the components of the generated wakefield in the THz plasma waveguide under consideration. For the numerical analysis, the plasma waveguide parameters were chosen such that the plasma wavelength in the region of the plasma background should be equal to $\lambda_{p(1)}=\lambda_{p(3)}=3\,mm$ (accordingly, the plasma frequency be $f_{p(1)}=99.9\,GHz$), and the eigenfrequency of the waveguide, resonant with the bunch, should be three times higher than $f_{p(1)}$. The parameters involved in the numerical analysis are presented in~Tab.\ref{Tabl_1}.
\begin{table}
\caption{\label{Tabl_1}Parameters of the plasma waveguide and the drive electron bunch.}
\begin{ruledtabular}
\begin{tabular}{cc}
Parameter&Value\\
\hline
Waveguide radius $R$&$0.6\,mm$\\
Inner tubular plasma radius $a$&$0.2\,mm$\\
Outer tubular plasma radius $b$& $0.4\,mm$\\
Plasma background density $n_{p(1)}, n_{p(3)}$&$1.23\cdot 10^{14}\,cm^{-3}$\\
Tubular plasma density $n_{p(2)}$&$1.63\cdot 10^{15}\,cm^{-3}$\\
Energy of bunch& $5\,GeV$\\
Charge of bunch $Q_b$&$1\,nC$\\
Length of bunch $L_b$&$0.2\,mm$\\
Radius of bunch $R_b$&$0.19\,mm$\\
\end{tabular}
\end{ruledtabular}
\end{table}
For solving the problem of electromagnetic field generation by a bunch of charged particles in a waveguide, it is essential to know its dispersion characteristics, which couple the eigenwave frequency to the longitudinal component of the wave number (at given parameters and geometry of the system), or, in other words, the phase velocity of the eigenwaves to the frequency and parameters of the waveguide. The waveguide dispersion under discussion is described by the equation $D(\omega ,k_z)=0$, and at arbitrary waveguide parameters values its general solution can be obtained only numerically. Figure~\ref{Fig:02} shows the TM-eigenwave dispersion curves (the first four modes), calculated with the dispersion equation $D(\omega ,k_z)=0$ for the above-given parameters of the waveguide and the drive bunch. It is pertinent to note some peculiarities of the given waveguide dispersion. The first peculiarity is that among TM-eigenwaves of the waveguide there are two slow waves (i.e., $s=1,2$). These are the low- and high-frequency surface waves. The second peculiarity lies in that for the chosen plasma waveguide parameters the low-frequency surface wave is forward, while the high-frequency surface wave is backward over the whole range of the wave number. In this case, the resonant frequencies are equal to 132.7 GHz and 298.7 GHz, respectively. The other TM-eigenwaves of the waveguide under study are the fast waves and cannot be excited by the relativistic electron bunch.
Aside from the dispersion dependence of the eigenmodes, analysis has also been done for the topography of the longitudinal electric field component, which is of interest in the consideration of charged particles interaction with the fields excited in the waveguides. That is, the necessity of analyzing the topography (transverse structure) of the electromagnetic field components stems from the need to know this electrodynamic characteristic when calculating the essential parameters that govern the acceleration efficiency, high-frequency power losses, etc. Figure~\ref{Fig:02} shows the topography of the longitudinal electric field component of the waveguide eigenmodes resonant ($k_z=\omega /v$) with the drive bunch, which were calculated with expressions~(\ref{eq:17})--(\ref{eq:20})). In the topography calculations, the terms $8Q_b/R_bL_b$ and $\Psi_{\parallel}(t,z,\omega_s)$ in the corresponding expressions were omitted. It can be seen from Fig.~\ref{Fig:02} that in the region of the plasma background at $r\leq a$ the transverse structure of the longitudinal Fourier electric field components is practically homogenous. This fact makes this mode favorable for using it as an operating accelerating mode from the standpoint of charged beam energy spread minimization. The axial electric field topography of the resonant mode of a high-frequency surface wave demonstrates that owing to the presence of the plasma background its radial homogeneity in the drive bunch region is distorted but little as opposed to the case of no background. This is what is illustrated by Fig.~\ref{Fig:02}. The reason is that the structure of the accelerating mode is mainly determined by the tubular plasma density, and the plasma background density is lower in comparison with it. It should be noted that in the absence of the plasma background ($n_{p(1)}= n_{p(3)}=0$) the resonant frequencies change to be 35.8 MHz and 291.5 GHz, respectively.

The resonant (Cherenkov) eigenwave frequencies of the waveguide are dependent on its parameters, in particular, on the plasma background and tubular plasma densities. Considering that we impose certain conditions on the frequency of the accelerating mode, we should know to which extent this mode  is responsive to changes in the plasma background and tubular plasma densities. Using the dispersion equation~(\ref{eq:12}), we have performed the corresponding numerical analysis for possible tolerances. The obtained results are presented in Fig.~\ref{eq:03}.
\begin{figure}
  \centering
  \includegraphics[width=0.49\textwidth]{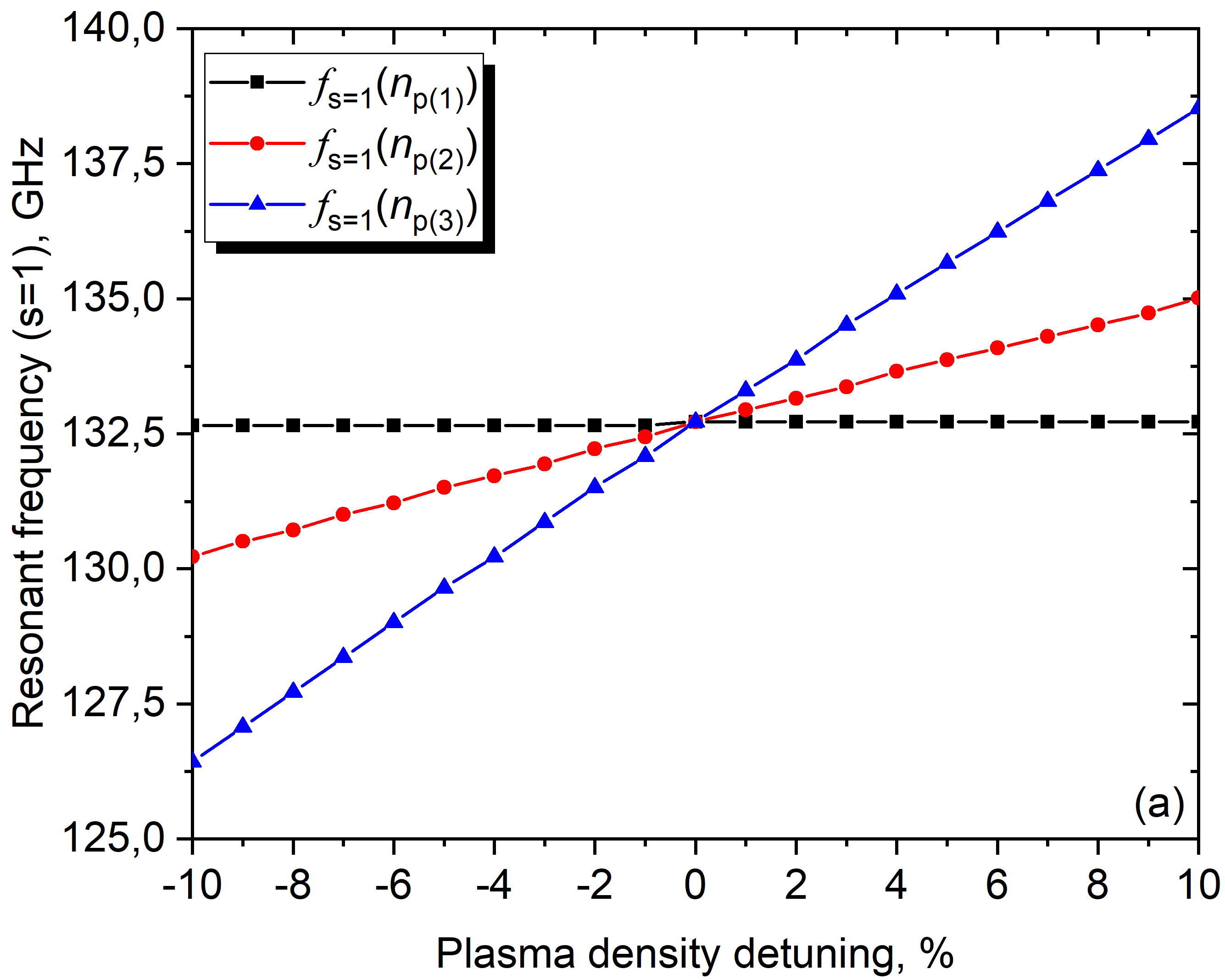}
  \includegraphics[width=0.49\textwidth]{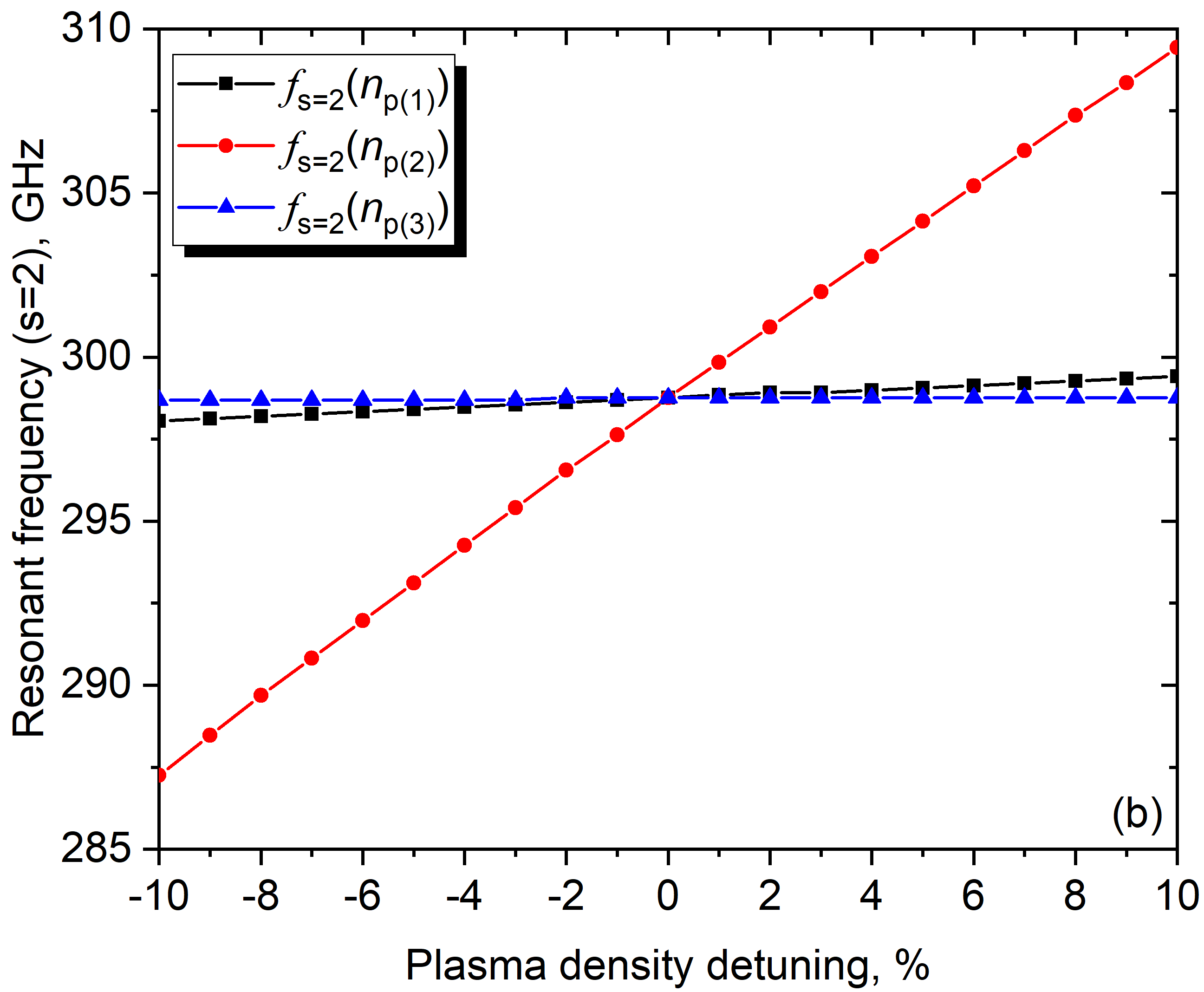}
  \caption{Resonant frequencies of (a) low-frequency and (b) high-frequency surface waves as functions of  density variations in one of the three plasmas, the other two plasma densities being unchanged.}\label{Fig:03}
\end{figure}
It can be seen that this function is linear for the both resonant frequencies. The resonant frequency value of the low-frequency surface wave appears most sensitive to variations in the density values of both the tubular plasma and the plasma background in the region close to the waveguide metal coating. Namely: (I) the change in the tubular plasma density within $\pm$10 \% results in the frequency value variation within $\pm$3 GHz ($\sim 2.3\%$), (II) the change of the plasma background density in the region close to the waveguide metal casing   within  $\pm$10 \% causes the frequency variation in the range of  $\pm$5 GHz ($\sim 3.8\%$). The resonant frequency value of the high-frequency surface wave is most sensitive to variations only in the tubular plasma density, and with its variations within $\pm$10 \% the frequency shift value attains $\pm$10 GHz ($\sim 3.3\%$). At that, the change in the plasma background density has no essential effect on the resonant frequency value of the high–frequency surface wave. That is, the frequencies of both the high- and low-frequency waves show even an appreciably weaker change than the plasma density change in both the tubular and background plasmas.

\begin{figure}
  \centering
  \includegraphics[width=0.49\textwidth]{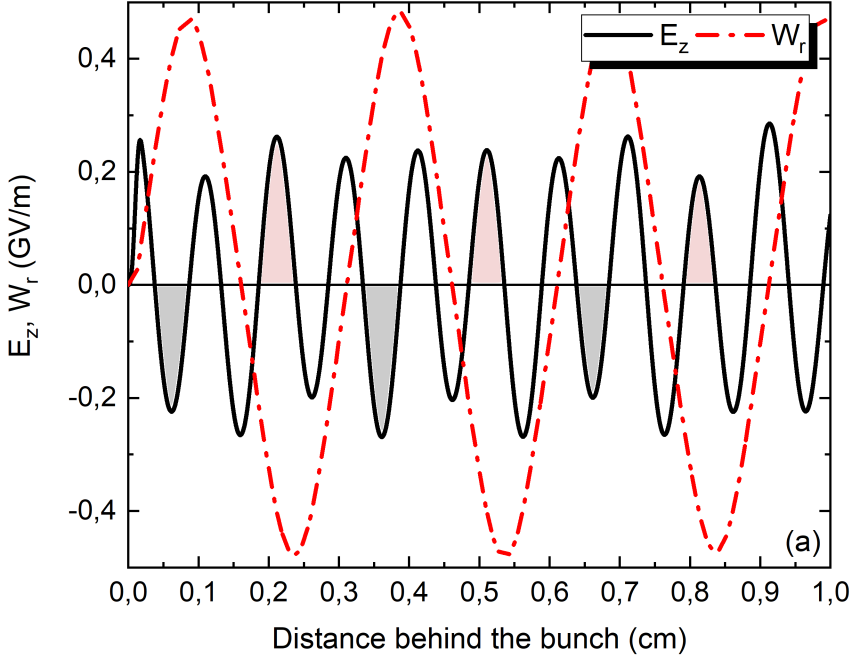}
  \includegraphics[width=0.49\textwidth]{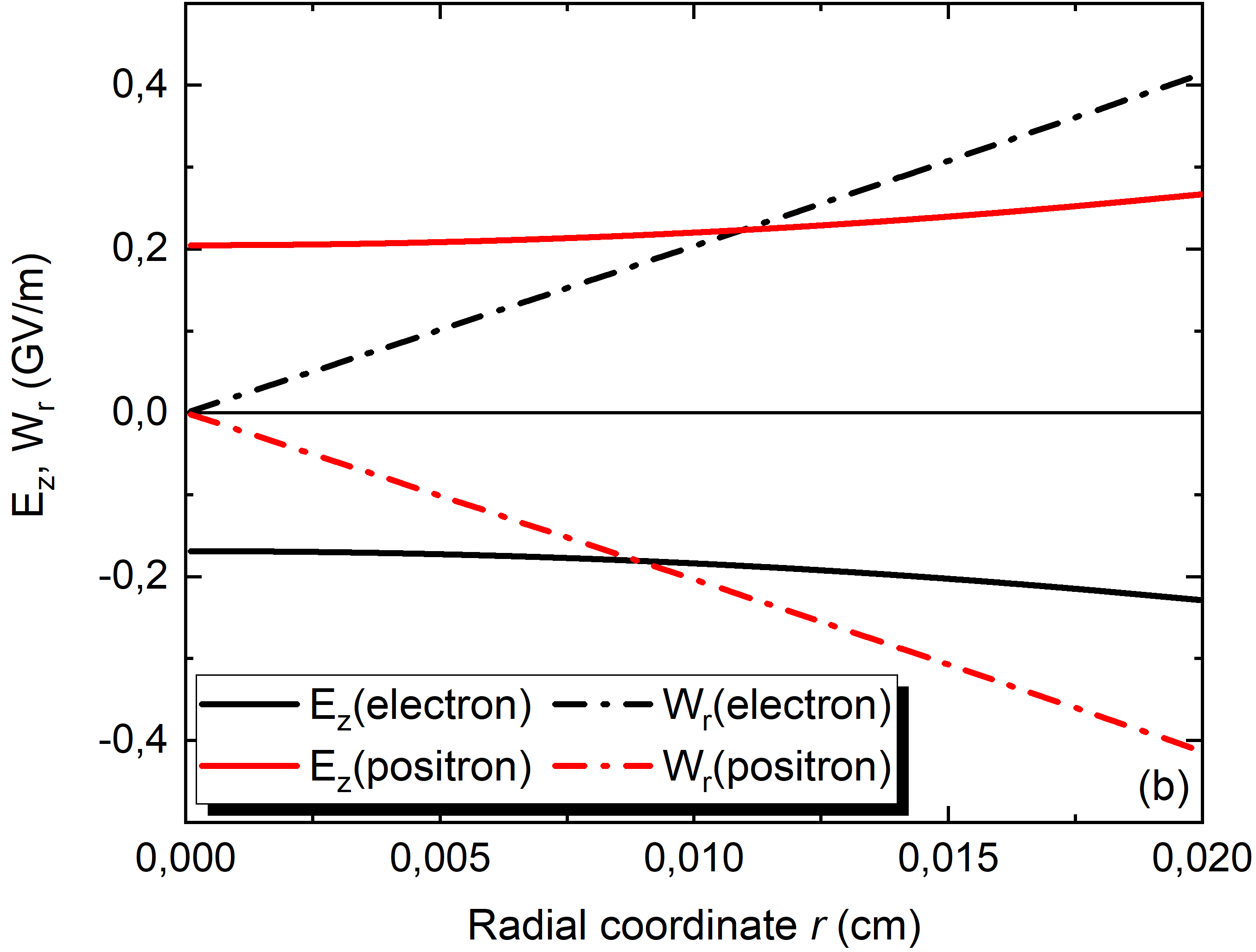}
  \caption{(a) Longitudinal distributions of axial $E_z$ and radial $W_r$ wakefields excited by the drive electron bunch at $r = R_b$. The bunch moves from right to left along the waveguide axis at a constant velocity. Possible positions of the test electron/positron bunches are shown as grey/pink regions under the plot of axial wakefield. (b) Transverse distributions of axial $E_z$ and radial $W_r$ wakefields in the range of radial coordinate values   $0\leq r\leq a$ for two values of the distance behind the drive bunch: 0.06 cm (test electron bunch position) and 0.21 cm (test positron bunch position).}\label{Fig:04}
\end{figure}

During electromagnetic field generation by the drive bunch and the test bunch acceleration in this field, the bunch particles are subjected to the longitudinal and transverse wakefields. The structure and amplitudes of these fields govern the space-time dynamics of the mentioned bunches, and, in particular, determine the dynamics of drive bunch energy loss as well as the dynamics of test bunch energy gain. Figure~\ref{Fig:04} shows the longitudinal distributions of the axial $E_z$ and radial wakefields $W_r=E_{r}-\beta H_{\phi}$, calculated at $r=R_b$ (i.e., on the drive bunch surface).
It can be seen, that the amplitude distributions of the both wakefields have clearly defined monochromatic character with slight shape distortions. The character of the longitudinal wakefield component distribution demonstrates the fact that behind the drive bunch there exist longitudinal coordinate regions, where both the electron and positron test bunches can be simultaneously accelerated and focused in the radial direction. Figure~\ref{Fig:04} also shows the transverse distributions of the axial $E_z$ and radial $W_r$ wakefields for two values of the distance behind the drive bunch. These are: 0.06 cm (that corresponds to one of the possible test electron bunch positions), and 0.21cm (that corresponds to one of the possible test bunch positions). In these cases, the character of the transverse distributions of $E_z$ and $W_r$ is the same (the only difference being in the sign). The $E_z$ distribution is slightly dependent on the radial coordinate, with insignificantly increasing from the waveguide axis towards the inner surface of the tubular plasma. This may not cause considerable increase of energy spread at acceleration of both the electron and positron test bunches.  The dependence of $W_r$ on the radial coordinate is linear. In the acceleration process, this will lead to that the peripheral part of the test bunch (both electron and positron ones) will be radially focused more strongly than its paraxial part. In this way, the radially stable acceleration of test bunches will take place, where the test bunches exhibit a partial radial compression, no filamentation and a low energy spread.

We now estimate the fulfillment of the plasma linearity condition for the numerical example under study. As is obvious from Fig.~\ref{Fig:04}, the transverse electric field component exceeds the longitudinal component. The maximum amplitude of the transverse field is attained at the boundary of the first background plasma of density $n_{p(1)}$ and the tubular plasma. The nonlinearity parameter for the excited wave is equal to $\epsilon = eW_r/mc\omega_{p(1)}$. Substituting there the maximum value from the data of Fig.~\ref{Fig:04} gives $\epsilon \sim 0.4$. That is, the background plasma in region (1) is weakly  nonlinear at the periphery of region (1). In the other cross sections of the waveguide the plasma linearity condition will be fulfilled still better.

\begin{figure}
  \centering
  \includegraphics[width=0.49\textwidth]{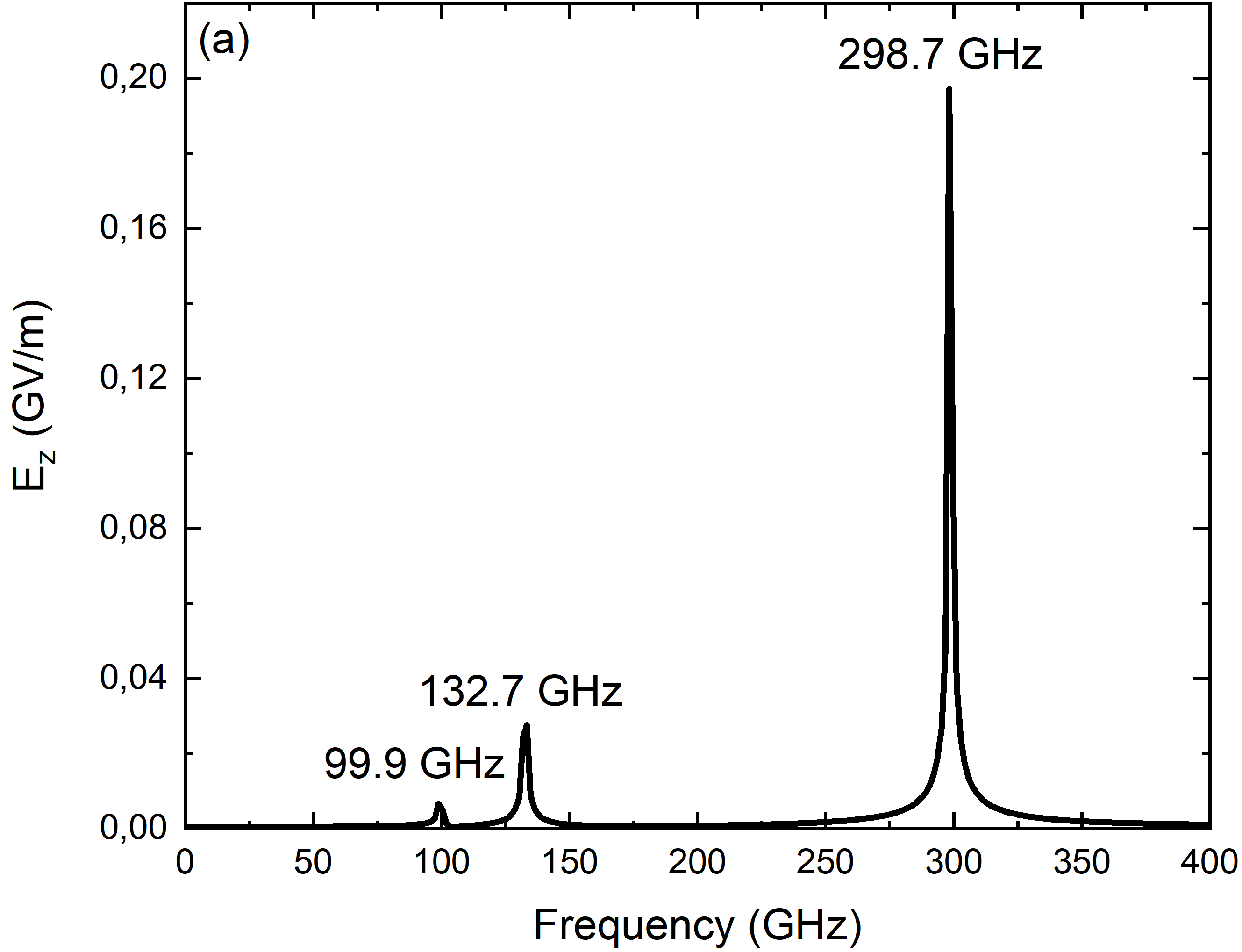}
  \includegraphics[width=0.49\textwidth]{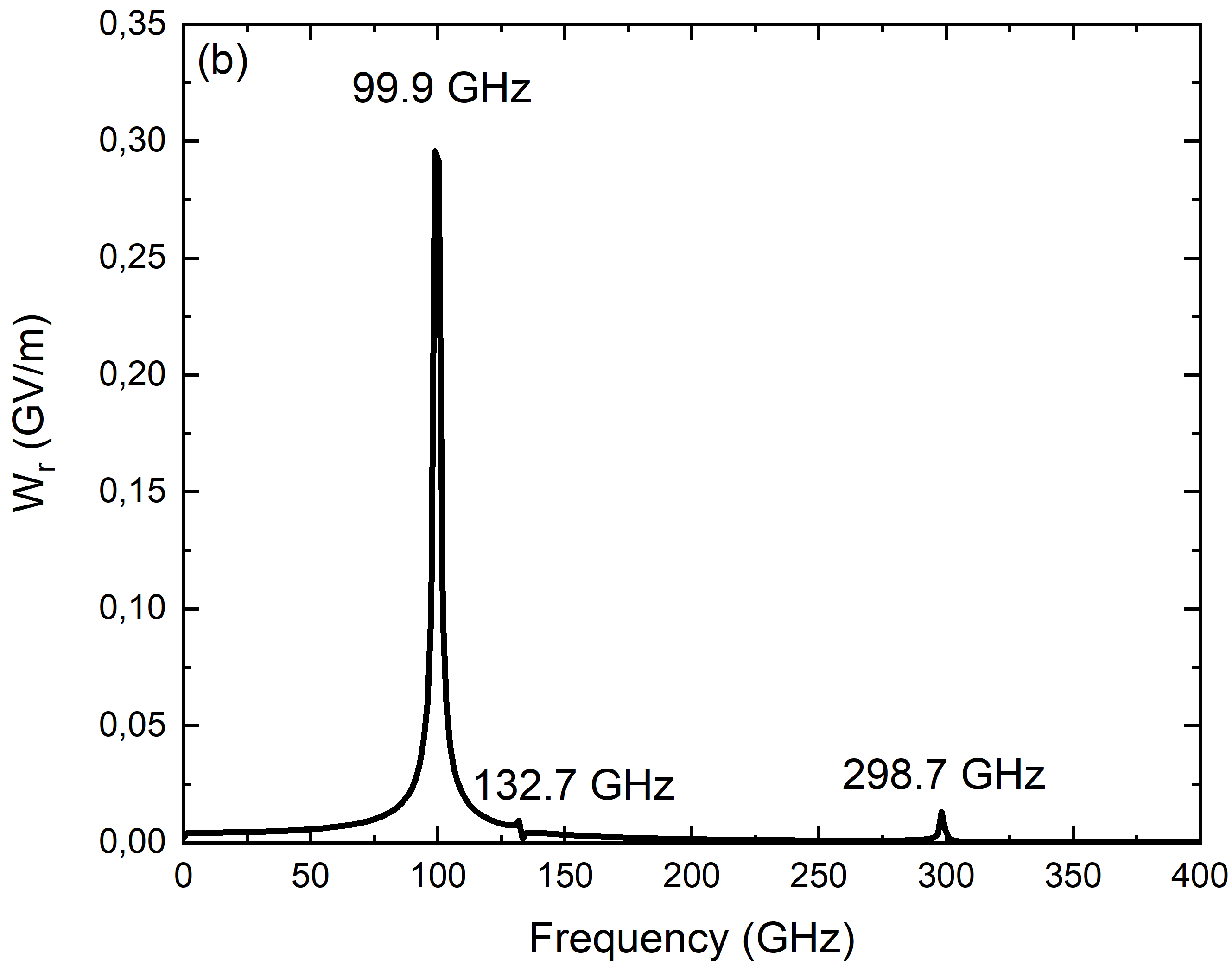}
  \caption{Spectra of the (a) axial $E_z$ and (b) radial $W_r$ wakefields excited by the drive bunch at $r=R_b$.}\label{Fig:05}
\end{figure}

Besides the analysis of spatial amplitude distributions, we have also made the spectral analysis of the excited longitudinal and transverse wakefields in the plasma channel region. The amplitude-frequency analysis is presented in Fig.~\ref{Fig:05}. The maximum in the spectrum of the axial wakefield $E_z$ corresponds to the frequency of the waveguide eigenmode resonant with the bunch  $f_{s=2}=298.7\,GHz$. It is evident that a slight distortion of the amplitude profile $E_z$ is due to the contribution from the axial components of the resonant waveguide eigenmode $f_{s=1}=132.7\,GHz$, and also from the plasma wave at the plasma frequency of the plasma background $f_{p(1)}=99.9\,GHz$. At the same time, the maximum in the spectrum of the radial wakefield $W_r$ corresponds to the plasma frequency of the plasma background $f_{p(1)}$, while the contribution from the radial component of the waveguide eigenmodes $f_{s=1,2}$, resonant with the bunch, is negligibly small.
Thus, it maybe concluded that in the plasma background region near the waveguide axis the generation of the longitudinal and transverse wakefields occurs at different frequencies. For purposes of acceleration, this means that it is possible to choose the injection time such that the test bunch (both electron and positron ones) will be at the maximum of the accelerating field, being simultaneously transversely focused.

\begin{figure}
  \centering
  \includegraphics[width=0.49\textwidth]{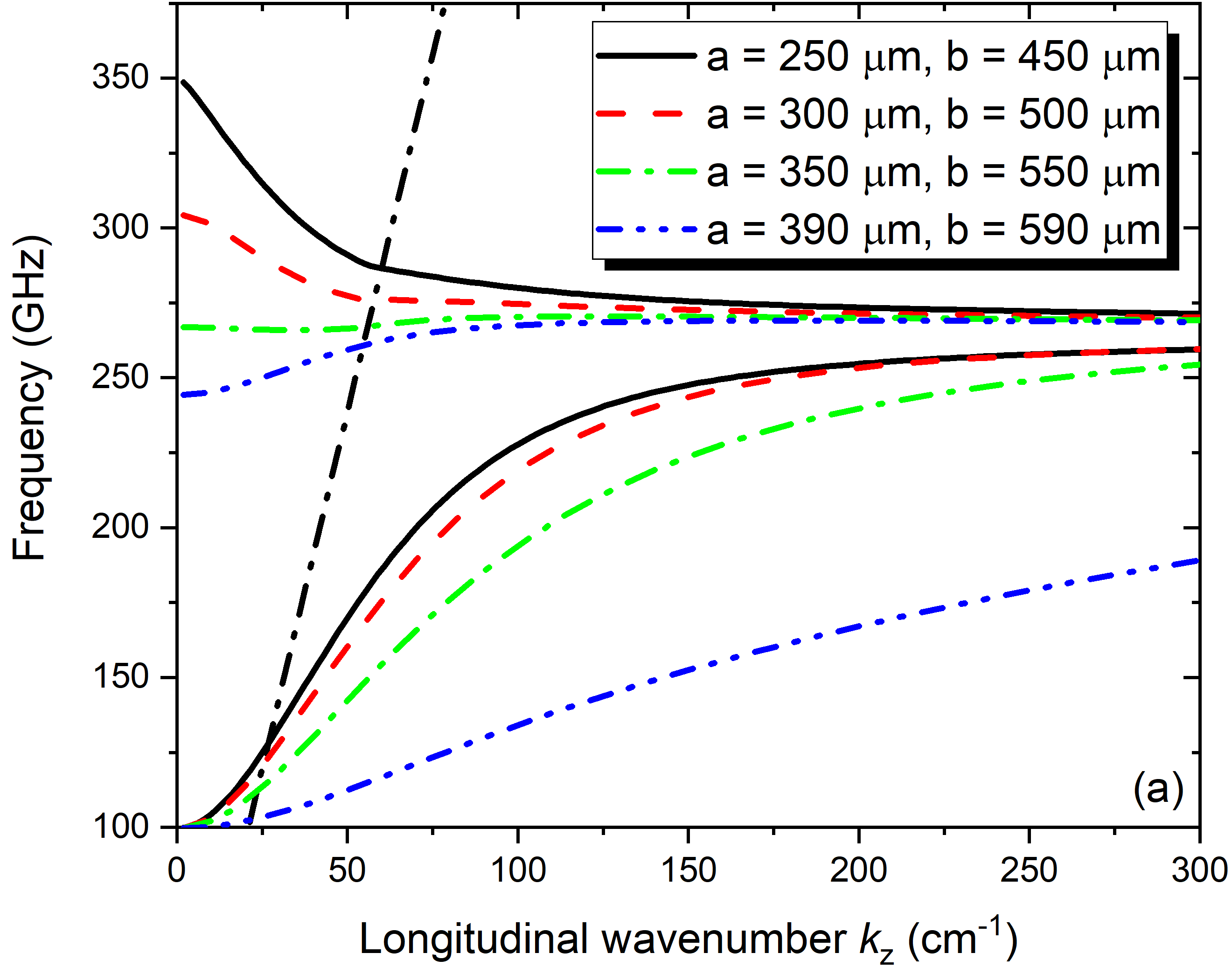}
  \includegraphics[width=0.49\textwidth]{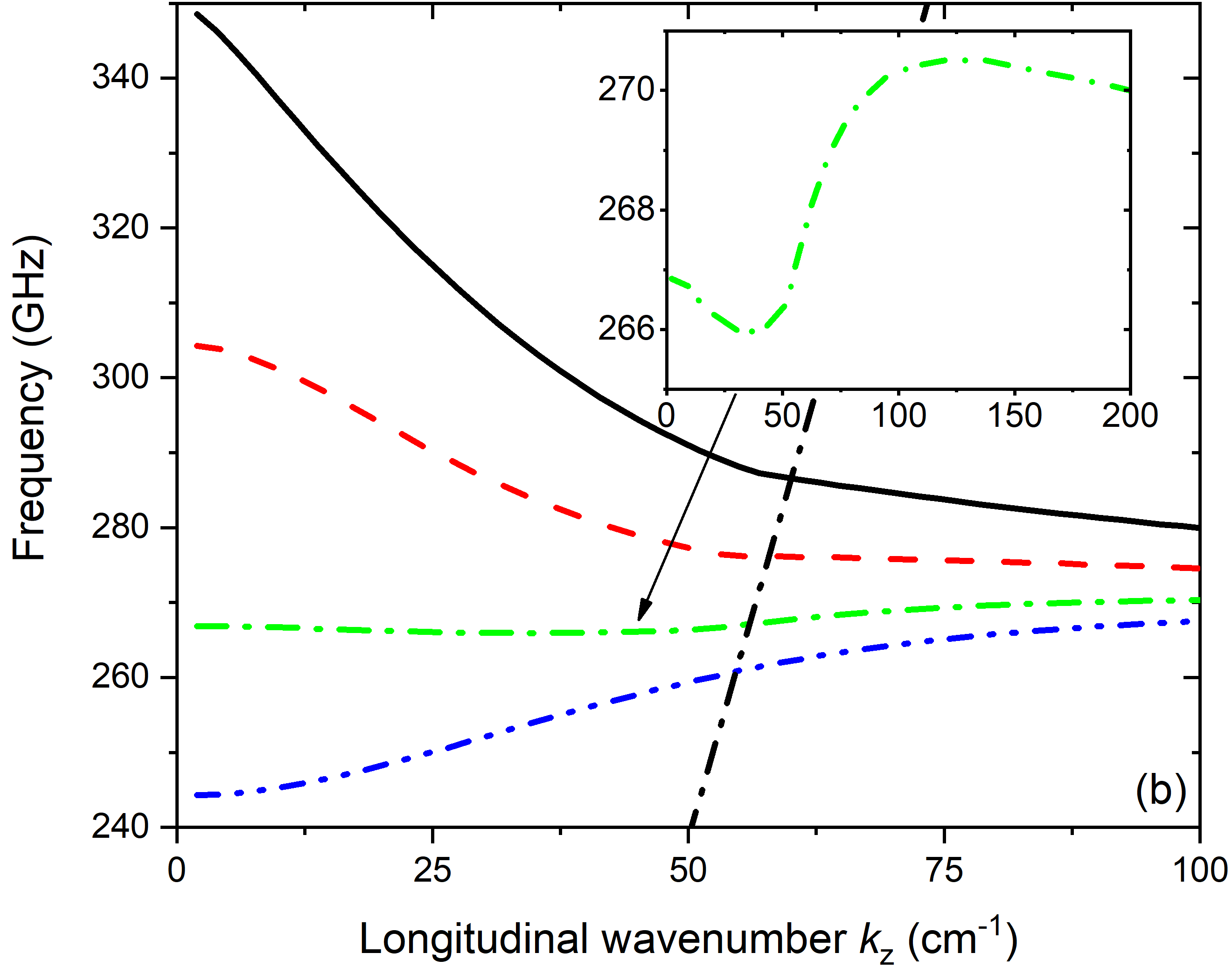}
  \caption{The dispersion of low-frequency and high-frequency surface waves (a), and a zoomed frequency range of the high-frequency surface wave dispersion (b), calculated for four values of the inner and outer radii of the tubular plasma at fixed thickness $b-a$, and density $n_{p(2)}$.}\label{Fig:06}
\end{figure}

The tubular plasma position in the waveguide relative to the plasma background can affect the generated wakefield. For fixed values of tubular plasma density and thickness, the space-time structure of the wakefield has been analyzed for different positions of the tubular plasma. First of all, the dispersion of the waveguide under consideration has been analyzed. The obtained results are presented in Fig.~\ref{Fig:06}.
 Figure~\ref{Fig:06} demonstrates that with change in the tubular plasma position there occur both the quantitative and qualitative changes in the dispersion dependence of the TM-eigenwaves. Namely, the tubular plasma withdrawal from the boundary of the drive bunch to the metal coating of the waveguide results in the following: (I) the resonant frequency values of both low-frequency and high-frequency surface waves get decreased, (II) the high-frequency surface wave is transformed from the backward wave with anomalous dispersion to the forward wave with normal dispersion. In this case, there exist the tubular plasma positions, at which the dispersion curves show the ranges of both anomalous and normal dispersion.

\begin{figure}[!th]
  \centering
  \includegraphics[width=0.32\textwidth]{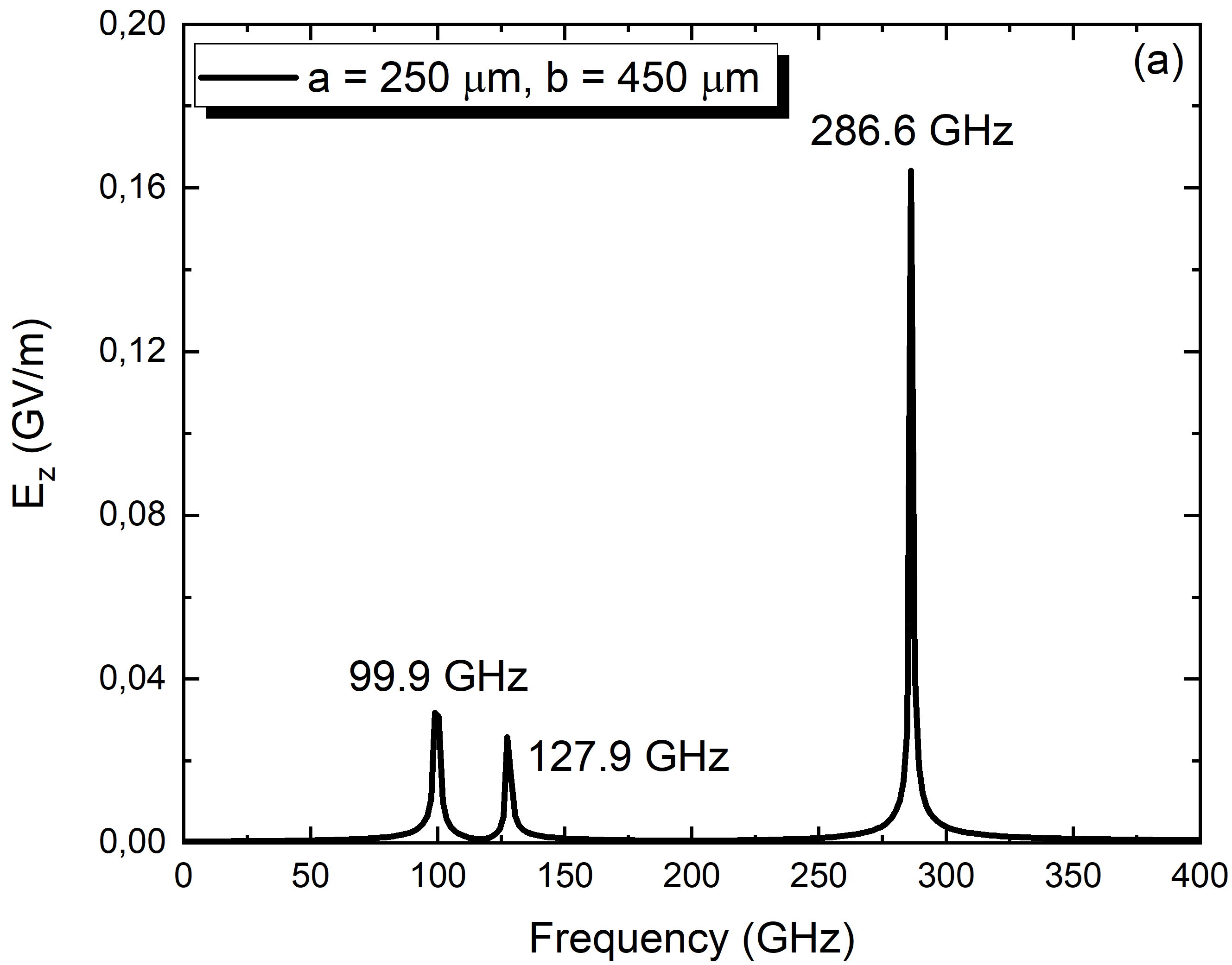}
  \includegraphics[width=0.32\textwidth]{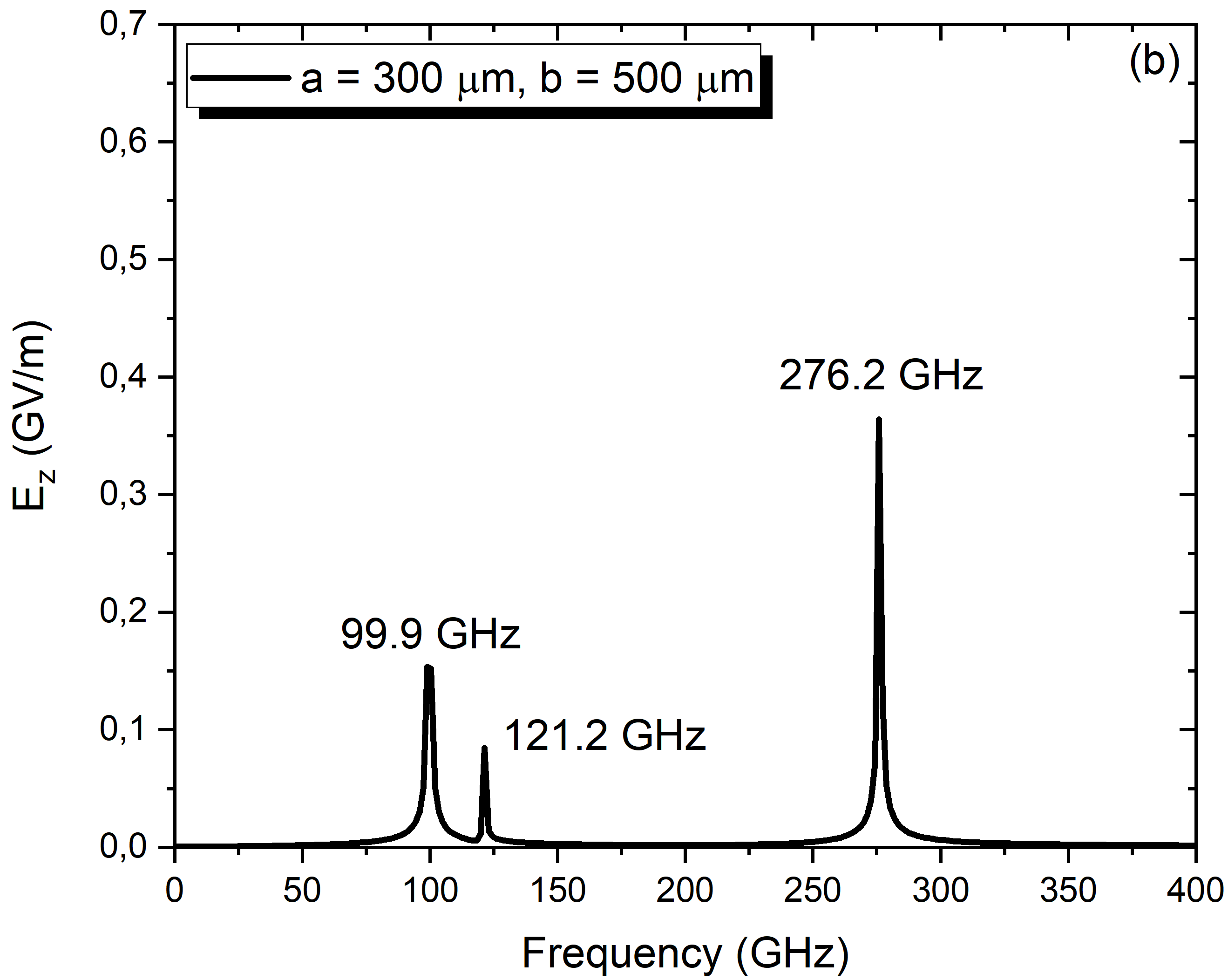}
  \includegraphics[width=0.32\textwidth]{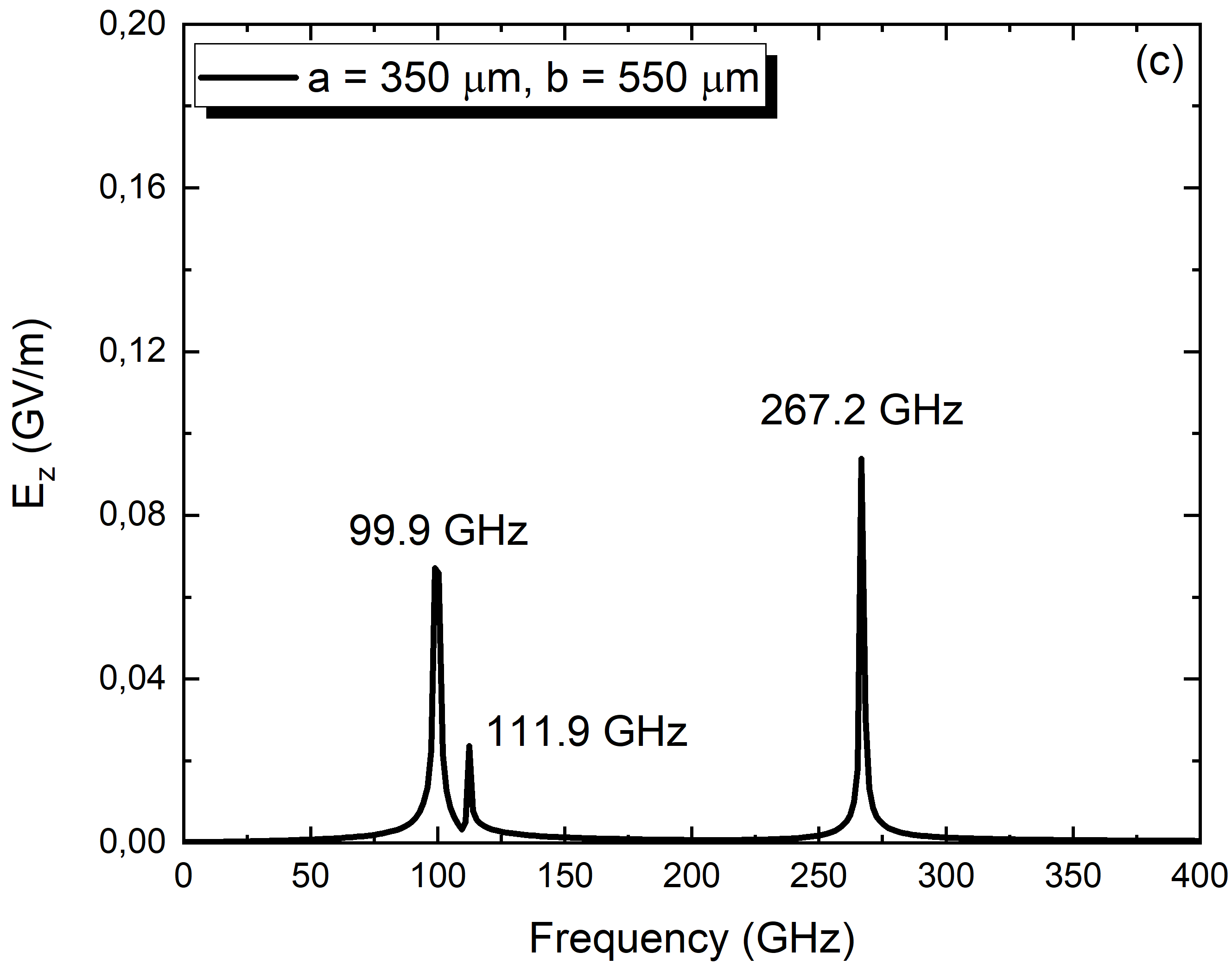}
  \caption{Spectra of the drive bunch-excited axial wakefield $E_z$, calculated at $r=R_b$ for three values of inner and outer radii of the tubular plasma ((a) $a=250\,\mu m, b=450\,\mu m$, (b) $a=300\,\mu m, b=500\,\mu m$, (c) $a=350\,\mu m, b=550\,\mu m$) at its fixed thickness and density.}\label{Fig:08}
\end{figure}

 Figure~\ref{Fig:06} shows the longitudinal distributions of generated axial and radial wakefields, calculated for three tubular plasma positions (at fixed plasma thickness and density values). From the figure we notice that the increase in the distance from the drive bunch surface to the inner surface of tubular plasma causes the longitudinal profile of the axial wakefield to take on the multiwave structure.
\begin{figure}
  \centering
  \includegraphics[width=0.32\textwidth]{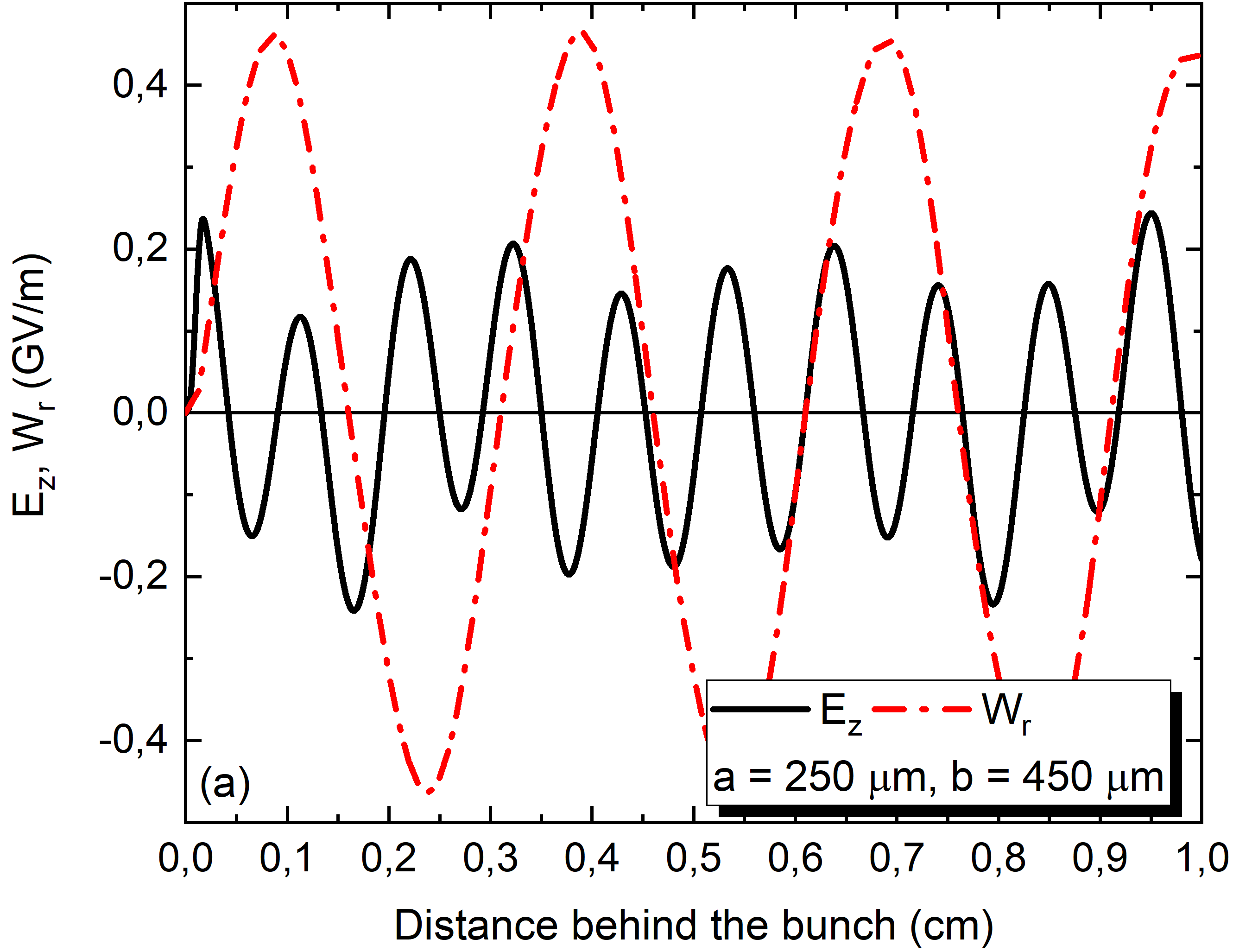}
  \includegraphics[width=0.32\textwidth]{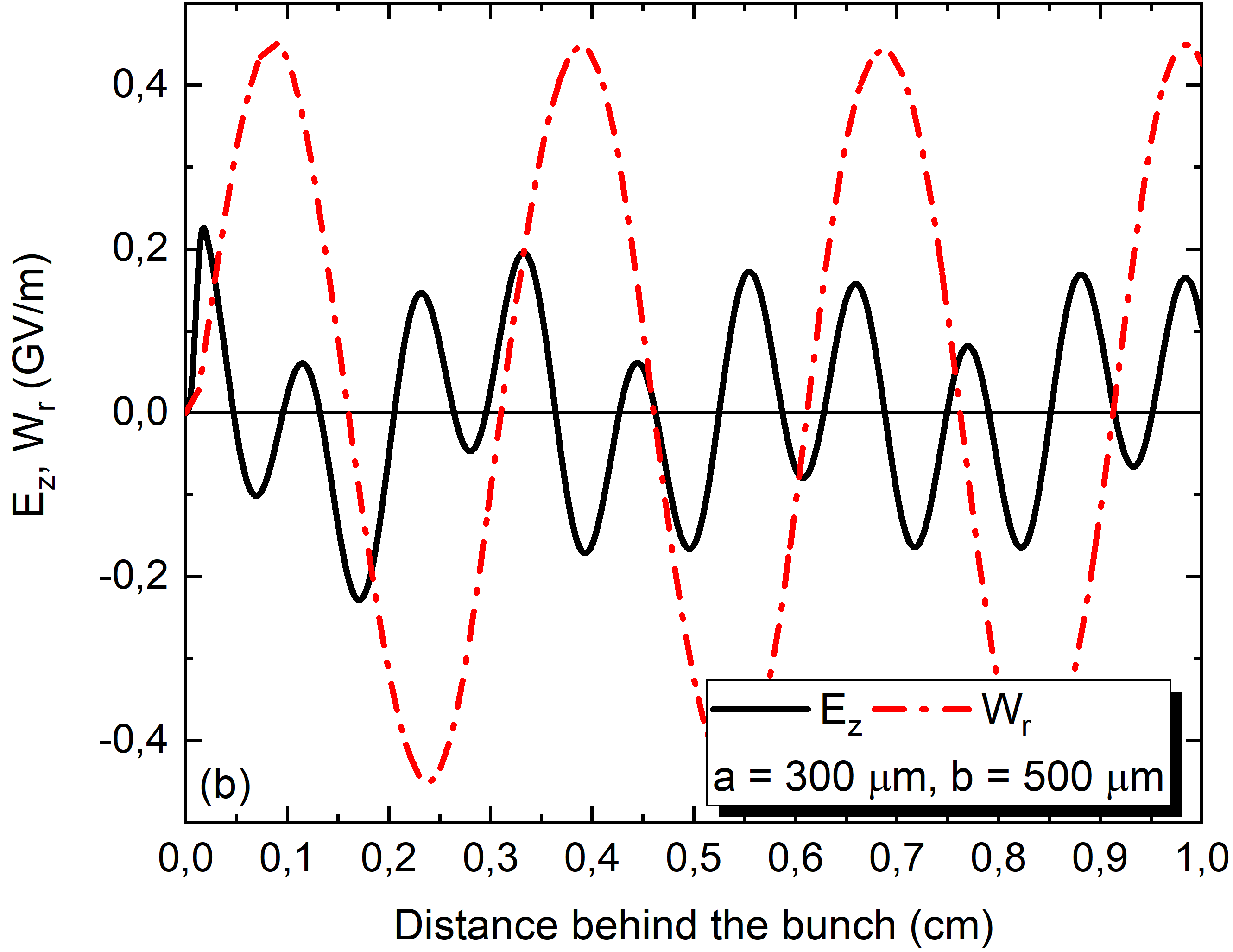}
  \includegraphics[width=0.32\textwidth]{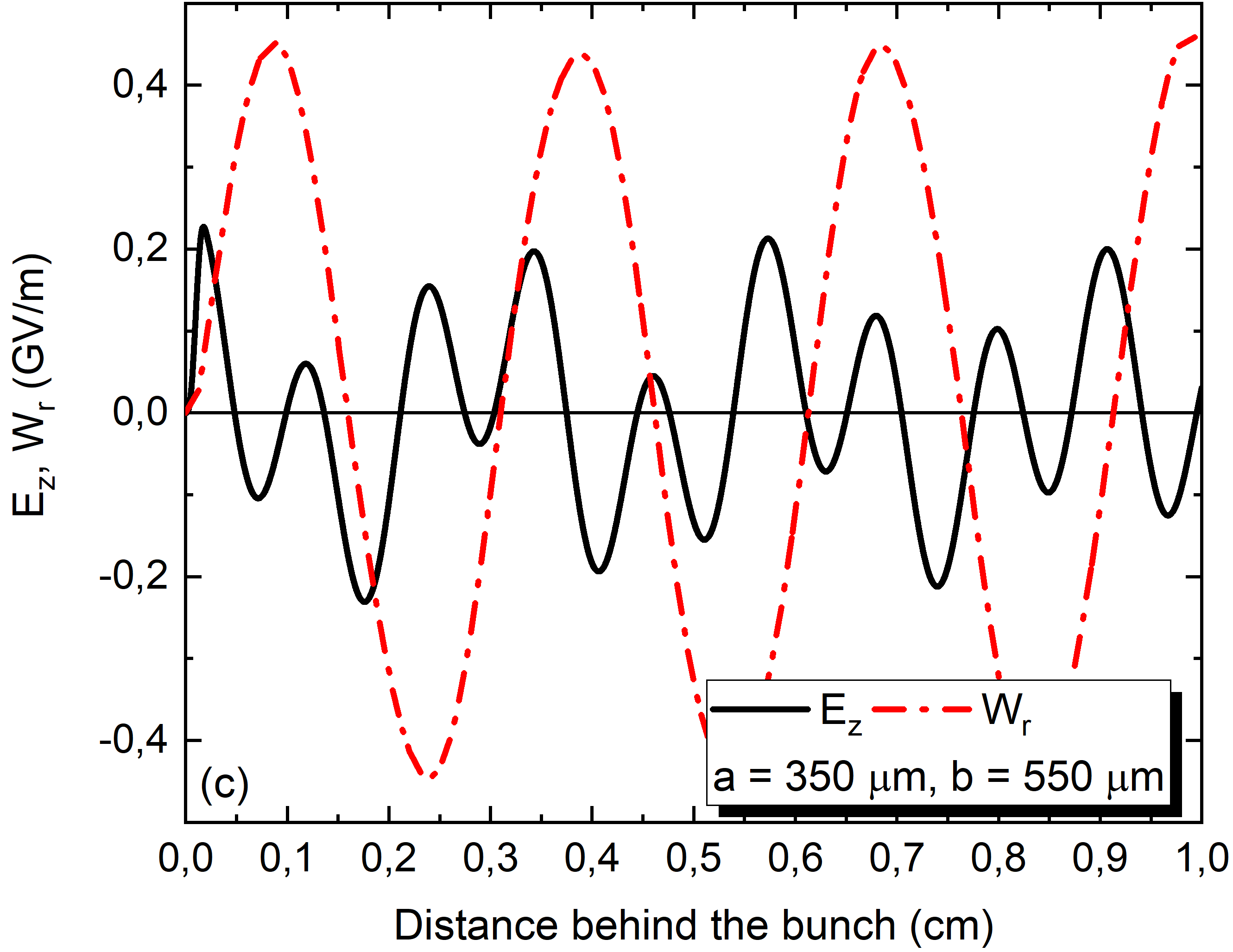}
  \caption{Longitudinal distributions of drive bunch-excited axial $E_z$ and radial $W_r$ wakefields calculated at $r=R_b$ for three values of internal and external radii of the tubular plasma ((a) $a=250\,\mu m, b=450\,\mu m$, (b) $a=300\,\mu m, b=500\,\mu m$, (c) $a=350\,\mu m, b=550\,\mu m$) at its fixed thickness and density values.  The bunch moves from right to left along the waveguide axis at a constant velocity.}\label{Fig:07}
\end{figure}
Besides, at that, there occurs the decrease in the axial component amplitude of the excited wakefield that acts on both the test electron and positron bunches.  As this takes place, no essential changes occur in the structure and amplitude of the transverse wakefield, because the latter is mainly contributed by the wakefield of the Langmuir wave. The change in the longitudinal structure of the excited axial wakefield should straightforwardly reflect on the change in its frequency spectrum and the mode structure. This is confirmed by the corresponding results of the undertaken frequency spectrum analysis (see Fig.~\ref{Fig:08}).
It follows from the results of the amplitude-frequency Fourier analysis that the change takes place in the relationship between the wakefield amplitudes of the plasma wave and the resonant frequency fields amplitudes. Namely, as the tubular plasma is moved away towards the waveguide metal coating, a decrease in the axial resonant frequency field amplitude occurs, whereas the plasma wave amplitude increases. As a result, the amplitude of the total axial wakefield is determined practically to the same extent by both the resonant frequency field and the plasma wave field, which have different wavelengths. This, in turn, leads to a multiwave profile structure shown in Fig.~\ref{Fig:07}.
\section{Conclusions}
In present paper it was developed a linear theory of wakefield excitation by the drive bunch in a cylindrical waveguide with transversely inhomogeneous plasma filling, which is represented by tubular plasma and the plasma background of different density. The integral Fourier expansion and the method of partial regions were used to construct closed form solutions for the excited radial and axial electric field components, and also, the azimuthal magnetic field component. The numerical analysis of the longitudinal and transverse amplitude distributions of the axial and radial wakefields has shown the feasibility of choosing such time values for injection of test electron/positron bunches, at which not only their acceleration would take place, but simultaneous transverse focusing as well. The obtained theoretical results can be used at preparation and design of the experiments to accelerate electron and positron bunches in plasma accelerating structures. In the future work, we intend to perform numerical PIC-simulations of the self-consistent dynamics of wakefield excitation, and to investigate the dynamics of both the drive and test bunches.
\section{Acknowledgments}
The study is supported by the National Research Foundation of Ukraine under the program “Excellent science in Ukraine” (project\# 2023.03/0182).
% If in two-column mode, this environment will change to single-column format so that long equations can be displayed.
%\begin{widetext}
%$$\mbox{put long equation here}$$
%\end{widetext}

% Figures should be put into the text as floats.
% Use the graphics or graphicx packages (distributed with LaTeX2e).
% Put the label that you will use with \ref{} command in the braces of the \label{} command.
%
% \begin{figure}
% \end{figure}

% \begin{table}
% \end{tabular}
% \end{table}

%\begin{acknowledgments}
% Put your acknowledgments here.
%\end{acknowledgments}

\bibliography{aipsamp}

\end{document}